\documentclass[useAMS,usenatbib,uselongtable]{mn2e}
\usepackage{natbib}
\usepackage{epsf}
\usepackage{subfigure}
\usepackage{subfig}
\usepackage{graphicx}
\usepackage{textcomp}
\usepackage{lscape}

\title[A comprehensive study of NGC 2023 with \textit{XMM-Newton} and \textit{Spitzer}]{A comprehensive study of 
NGC 2023 with \textit{XMM-Newton} and \textit{Spitzer}\thanks{This 
   publication makes use of 
   data products from the Two Micron All Sky Survey, which is a joint project of 
   the University of Massachusetts and the Infrared Processing and Analysis 
   Center/California Institute of Technology, funded by the National Aeronautics 
   and Space Administration and the National Science Foundation. 
   This research has also made use of the Centre de DonnŽes Astronomiques de 
   Strasbourg (CDS) tool Aladin.}}
\author[M. A. L\'opez-Garc\'ia et al.]
{M. A. L\'opez-Garc\'ia$^{1}$\thanks{E-mail: mal@astrax.fis.ucm.es}, 
 J. L\'opez-Santiago$^{1}$, 
 J. F. Albacete-Colombo$^{2}$, 
 \newauthor P. G. P\'erez-Gonz\'alez$^{1}$$^{,}$$^{3}$ 
 and E. de Castro$^{1}$\\
$^{1}$Departamento de Astrof\'{\i}sica y Ciencias de la Atm\'osfera, Universidad Complutense de Madrid,
              E-28040 Madrid, Spain\\
$^{2}$Centro Universitario Regional Zona Atl\'antica (CURZA) 
             Universidad Nacional del COMAHUE, Monse\~nor Esandi 
             y Ayacucho (8500),  \\Viedma (Rio Negro), Argentina.\\
$^{3}$Associate Astronomer at Stward Observatory, The University of Arizona.}
\begin{document}

\date{Accepted .... Received ...}

\pagerange{\pageref{firstpage}--\pageref{lastpage}} \pubyear{2002}

\maketitle

\label{firstpage}

\begin{abstract}
 
Nearby star-forming regions are ideal laboratories to study high-energy emission of different stellar
populations, from very massive stars to brown dwarfs.
NGC~2023 is a reflection nebula situated to the south of the Flame Nebula (NGC~2024) and at the 
edge of the H~\textsc{ii} region IC~434, which also contains the Horsehead Nebula (Barnard~33). NGC~2023, 
NGC~2024, Barnard~33 and the surroundings of the O-type supergiant star $\zeta$~Ori constitute the south 
part of the Orion B molecular complex. %, a nearby giant molecular cloud were star formation occurs actively.
In this work, we present a comprehensive study of X-ray emitters in the region of NGC~2023 and its surroundings. 
We combine optical and infrared data to determine physical properties (mass, temperature, luminosity, 
presence of accretion disks) of the stars detected in an {\it XMM-Newton} observation. This study has allowed us 
to analyze spectral energy distribution of these stars for the first time and determine their evolutionary stage. 
Properties of the X-ray emitting plasma of these stars are compared to those 
found in other nearby star-forming regions.  
The results indicate that the stars that are being formed in this region have characteristics, 
in terms of physical properties and luminosity function, similar to those found in the 
Taurus-Auriga molecular complex.

%, more similar to those found in the Taurus-Auriga 
%molecular complex, than of its neighbor Orion Nebula Cluster (ONC).
%
%We remark the detection of three class I objects (6\% of the X-ray detections) that seem to be
%underluminous in X-rays when compared to class 0/I objects in the ONC. 

\end{abstract}


\section{Introduction}
\label{intro}

Star-forming regions are the best laboratories to study the different physical processes that 
give rise to X-ray emission. X-ray photons are little absorbed by interstellar material
and young stars show higher levels of X-ray emission than main-sequence stars. 
%X-ray photons, in particular those with higher energy, scape the molecular cloud and reveal the stellar source. 
%
High-energy radiation plays an important role in the development and evolution of stellar protoplanetary 
disks \citep{bal00, sca01} and planetary atmospheres \citep[e.g.][]{pen07} and contributes to the disruption 
of star formation by causing molecular material blow out. Outflows, jets and winds are the main 
mechanisms to modify the molecular cloud environment, except for supernova explosions. They displace 
cloud material but also trigger star formation. 

%X-ray emission in young late-type stars is produced by hot plasma trapped in the stellar corona.  
%The X-ray emission level of these stars depends on their rotation rate \citep{piz03}. 
%
%Hence, when a classical T Tauri star (CTTS) looses its accretion disk and spins up, becoming a weak-line 
%T Tauri star (WTTS), its X-ray emission level increases \citep{tel07}. Later, when the star spins down again 
%due to the loose of angular momentum during the pre-main sequence and main-sequence phases, its X-ray 
%emission level decreases again. 
%
%What it is still not 
%well-known is when X-ray emission starts \citep{sci08}. Recent studies show that class I objects 
%(protostars with an accretion disk and a circumstellar envelope) do emit in X-rays \citep{tsu05,gia07b,pri08,pil10}.
%Those X-ray detected protostars show some X-ray emission characteristics that are different from those 
%observed in CTTS or WTTS, such as the presence of the iron fluorescence line at 6.4 keV. However, this
%line has been also detected during flaring event in more evolved stars \citep[e.g.][]{tes08,lop10b}. 
%More detailed studies must be carried out to progress in the understanding of the appearance of  
%powerful magnetic fields in late-type stars.}

The Orion giant molecular complex (see Fig.~\ref{comap}) consists of two main large structures, usually
referred to as Orion A and B, and a number of less massive filaments 
\citep[see][for a detailed description]{gen89}. The Orion A molecular cloud contains the 
Orion Nebula Cluster (ONC) which is the closest OB association to the sun 
($\sim 450$ pc). Because of its proximity, the ONC is one of the most visited targets for 
numerous studies, including the relation between cloud kinematics and star formation
\citep[see][and references therein]{har07}, protostellar and circumstellar disk 
characteristics \citep[e.g.][]{fan09, ing09, wil11} and the properties of the X-ray emission
of young stellar objects and T Tauri stars \citep[see][and the rest of the articles of the 
Chandra Orion Ultradeep Project (COUP) 
in the special issue of the ApJS]{pre05}. An extensive study of Orion A combining data 
from the \textit{XMM-Newton} and \textit{Spitzer} missions is just being carried out 
\citep[see preliminary results in][]{wol10,pil10}. 

Equally interesting for studies on stellar formation is the Orion B molecular cloud. The complex 
extends roughly south-north and contains the Horsehead Nebula, NGC 2023, NGC 2024, 
and several other reflection nebulae, such as NGC 2068 and NGC 2071. The southern part of 
Orion B (L~1630) borders the large H\,\textsc{ii} region IC~434, which is expanding into the molecular 
cloud. The interface between the molecular cloud and the H\,\textsc{ii} region (IC~434) is seen 
as a bright ridge of glowing gas with the Horsehead Nebula and several smaller pillars \citep{moo09}. 
The Horsehead Nebula points directly towards the multiple system $\sigma$ Ori, 
which is ionizing IC 434. 
At a distance of 350--450 pc, the Horsehead Nebula is the closest pillar to the Sun and represents an ideal 
laboratory to study the emergence and evolution of X-ray emission in very young stars, their X-ray properties 
and the influence of X-ray emission and its variability on the heating and evolution of proto-planetary disks
\citep[see infrared studies by][]{bow09,moo09}.

NGC 2023 is located about 15 arcmin to the northeast of the pillar. In contrast to NGC 2024, that contains 
an embedded stellar cluster with more than 200 members and several protostar candidates 
\citep{hai00,ski03}, NGC 2023 shows low stellar density. Some radio clumps have been detected 
using the VLA \citep{ang02,rei04}. One of them, the source NGC 2023 MM 1, is a confirmed
(very cold) Class 0 source driving a large and well collimated molecular outflow \citep{san99}.
Three Herbig-Haro objects were identified by \citet{mal87}: [MOW87] HH1, HH2, and HH3. 
\citet{moo09} associated HH3 with the classical T Tauri star V615 Ori, but the association of
HH1 and HH2 with a detected source is more uncertain. The authors associated HH2 to the 
mid-infrared source MIR-51, while HH1 could be associated to the radio source NGC 2023 
MM3, the mid-infrared source MIR-62 or even the near-infrared source NIR-15, merely at 40
arcsec from HH1. %Hence, NGC 2023 represents a good example} of star-forming region 
%that is being modeled by high-energy processes. 

The age of NGC~2023 is not well established. \citet{alc00} derived ages in the range 
$0.5 < \tau_\mathrm{age} < 7.0$ Myr for stars in the L~1641 and L~1630 clouds with an X-ray 
counterpart. For NGC~2024, \citet{eis03} determined an age of 0.3 Myr from the disk frequency, 
while \citet{ali98} proposed 0.5 Myr as the age of this cluster, based in infrared photometry. 
The coexistence of both radio clumps and stars without a protoplanetary disk in the region of 
NGC~2023 \citep[e.g.][]{moo09} suggests that star formation has taken place during 
several million years, although this hypothesis should be tested robustly.

In this work, we present for the first time a comprehensive study of the X-ray sources in the 
star-forming region NGC 2023 and its surroundings. We give general X-ray emission 
characteristics of the sources (temperature, column density and emission measure) and 
compare them with the results obtained for the nearby regions %NGC 2024 \citep{ski03}
Taurus \citep{gudel2007} and Orion Nebula Cluster \citep[ONC; ][]{pre05}. We complete a multi-wavelength study of the 
X-ray detected sources using data from the literature. In addition, archival observations from the 
\textit{Spitzer} infrared observatory are analyzed. We then classify the X-ray sources into classes I, II 
and III objects according to their infrared characteristics. %All this analysis has allowed us to achieve 
%some conclusions on the possible influence of the environment in the star formation process 
%of this region. 

%_____________________________________________________________
\begin{table}
  \caption{NGC 2023 \textit{XMM-Newton} observation details on each EPIC detector.} 
  \label{tab1}              
  \tiny
  \begin{tabular}{@{}c c c c}      
    \hline             
     & PN & MOS1 & MOS2 \\
    \hline                        
    Exp. time (ks)& 25.09 & 29.13 & 29.12\\
    Effective exp. time (ks) & 13.60 & 15.00 & 15.80\\
    Filter & Medium & Thin & Medium\\
    Obs. mode & Extented Full Frame & Full Frame & Full Frame\\
    \hline                                   
  \end{tabular}
\end{table}
%_____________________________________________________________

%_____________________________________________________________
 \begin{figure}
 \centering
 \includegraphics[width=8.5cm]{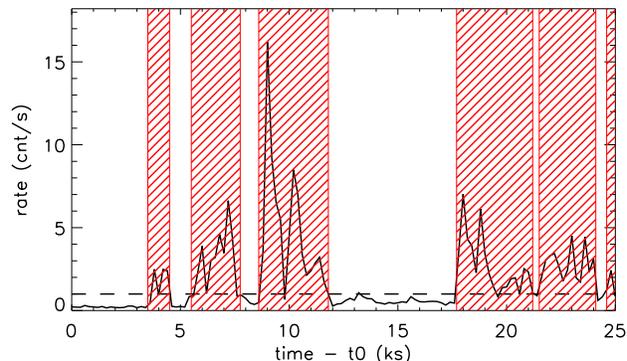}
   \caption{Lightcurve of the whole EPIC-PN detector at $E > 10$ keV. The dashed
   line indicates the recommended value for excluding flaring high background by the 
   SAS manual. Shaded regions are those showing high background flaring and, 
   therefore, are those discarded for our analysis.}
   \label{fig2}
 \end{figure}
%_____________________________________________________________

\section{X-ray observation and data reduction}
\label{obs}

The \textit{XMM-Newton} observation of the NGC 2023 reflection nebula (ID 0112640201)
was part of an observing program aimed at studying deeply embedded stellar clusters in Orion B. 
The observation was performed as a single, 30 ks long exposure on 2002 during satellite  revolution 
419. The EPIC (European Photon Imaging Camera) was used in full-frame mode as primary instrument.
The observation was centered at $\alpha = 05\mathrm{h}41\mathrm{m}47.20\mathrm{s}$ and 
$\delta = 02\mathrm{d}16\mathrm{m}37.0\mathrm{s}$. Table~\ref{tab1} summarizes the 
information of the parameters used for each EPIC detector.

The data reduction was done through the \textit{XMM-Newton} Science Analysis System (SAS) software, 
version 10.0. Images and spectra were produced with the standard SAS tools. We rejected high 
background periods from the analysis of the lightcurve of the whole detector
(independently for EPIC-MOS and PN) and selected time periods with constant, low X-ray emission at 
$E > 10$ keV (Fig.~\ref{fig2}). Bad events and noise were also 
rejected to create the final (GTIs; good time intervals) event tables. From our analysis, we conclude that
the observation was largely contaminated by high X-ray emission background periods (usually attributed 
to protons and other relativistic particles produced during solar flares). After our processing, 
the useful exposure time was reduced to 13.6 ks for PN, 15 ks for MOS1 and 15.8 ks for MOS2. 
Hereafter, we use the cleaned event lists.

\begin{figure}
 \centering
 \includegraphics[width=8cm]{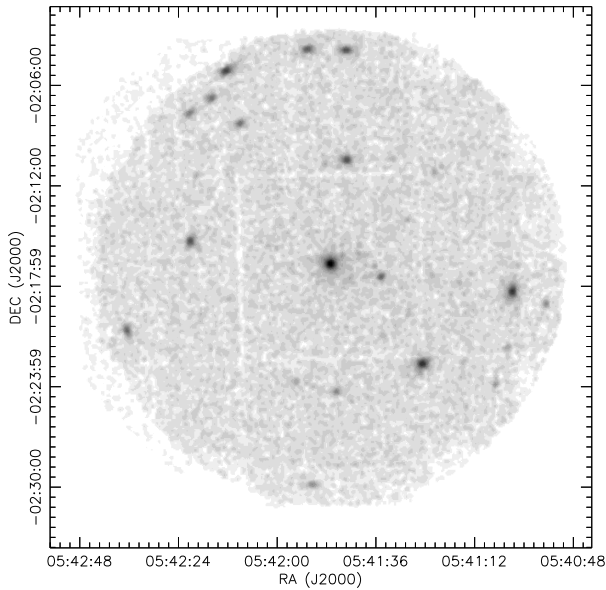}
   \caption{Composite EPIC (MOS+PN) image of the \textit{XMM-Newton} FoV 
                 in the 0.3--8.0 keV energy band.}
   \label{fig1}
 \end{figure}

In Fig.~\ref{fig1}, we show a mosaic of PN and MOS images of NGC 2023 in the energy 
band 0.3--8.0 keV, created with the task \textit{emosaic}.
We used the SAS task \textit{edetect\_chain} to reveal sources at three different 
energy bands (0.3--1.2, 1.2--2.5, 2.5--4.5 keV) in PN, MOS1 and MOS2 separately. 
A second stage consisted of using the 0.3-8.0 keV energy band
to confirm the first detections.  A total of 50 sources were detected. 
After a careful inspection that includes a cross-match with 
optical and infrared catalogs (see Section~\ref{cross}), we discarded 14 of them 
because they were multiple or spurious detections, typically at the detector border
or close to a bright source.
%The flux was determined using a conversion factor (CF) between count-rate 
%and flux, CF= 2.87$\times$10$^{-12}$ erg$\cdot$ph$^{-1}$. This conversion factor was obtained 
%from spectral fitting of the good count-rate object of the observation. 
The list of the X-ray sources is given in appendix (Table~\ref{tabX}). We have removed those 
sources considered as spurious detections during the optical and infrared analyses (Section~\ref{cross})
from the tables, although we have maintained the original numeration  for convenience.
X-ray sources of \citet{yam00} are identified in the table last column as [YKK2000] followed by 
the ID number given by the authors.  

%_____________________________________________________________

\section{Infrared properties of X-ray sources}
\label{cross}

\subsection{Spitzer}
\label{spitzer}

The region of NGC 2023 and its surroundings present a large amount of gas and dust 
that produces high optical and near-infrared absorption \citep[see][]{bow09}. The absorption 
by the gas and dust is less dramatic in the mid-infrared, in particular in the \textit{Spitzer-IRAC} 
bands and in radio-wavelengths. \citet{moo09} used archive \textit{Spitzer} observations of 
this region and sub-millimetric observations performed by themselves with the Submillimeter
Common User Bolometric Array \citep[\textit{SCUBA};][]{hol99} to create an infrared census of young 
stellar objects and protostars.  \citet{moo09} analyzed only part of the \textit{Spitzer}-IRAC observations,
close to the Horsehead. The region they selected overlaps only partially with the \textit{XMM-Newton} 
observation, that is centered on NGC 2023.
We have reanalyzed the \textit{Spitzer} data covering the \textit{XMM-Newton} field of view to look 
for infrared counterparts of all the X-ray detected sources.

For our work, 
we used the observations with \textsc{aorkey} numbers 8773120 and 8773632. Two collections
of images are present in those observations for each IRAC channel, corresponding to two different 
configurations with frame-time 0.6 and 10.4 seconds, respectively. We treated both datasets separately.
The regular IRAC pipeline of MOPEX (version 18.4.9) was used for the reduction and the mosaic 
creation. At the end of the reduction process, we had two mosaics for each IRAC channel and 
observation, for a total of four mosaics per channel. Photometric analysis was performed in each 
mosaic independently. 

It was not the aim of this work to obtain IRAC fluxes for all the sources in the field. We 
determined infrared fluxes only for the IR counterparts of the X-ray sources
(see Fig.~\ref{fA1}). To derive the fluxes, we used the \textsc{daophot} package included 
in IDL, which is an adaptation of the Fortran code also used in IRAF\footnote{IRAF: http://iraf.noao.edu/}. 
Aperture photometry with three different extraction radii (2, 3 and 4 arcsec) was performed. 
For the background (sky) subtraction, we used an annulus with inner and outer radii 20 and 30 arcsec, 
respectively. The mosaics performed by MOPEX are in units of MJy/sr. We transformed them to mJy 
and applied an aperture correction. Aperture corrections were estimated from semi-empirical 
point response functions (PRFs)\footnote{``Spitzer pixels are relatively big relative to point sources, 
so MOPEX fits with Point Response Functions (PRFs), which take into account intra-pixel sensitivities.''}
given in the Spitzer Manual. Their values are given in Table~\ref{tab2} for each channel. 
The uncertainties include the effects of typical World Coordinate System (WCS) random alignment 
errors \citep[always less than 1 pixel; see][]{per08}. For each source, we determined magnitudes
in AB system and Vega system. The latter were determined using the flux to magnitude conversion 
of the GLIMPSE\footnote{GLIMPSE: http://www.astro.wisc.edu/sirtf/}.
%
%Values determined for the fluxes using different aperture radii were very similar. 
For each infrared source, 
fluxes determined with extraction radii of two, three and four arcsec are very similar except 
for the two most brilliant IRAC sources, that show differences of up to 10\% in the four channels. 
The same statement is applicable to the comparison between observations with frame 
times 0.6 and 12 seconds. In the latter case, the differences arise from the 
fact that those sources are very close to the saturation regime for long exposures. 
To avoid problems with those sources, we decided to use only the 
images with shorter exposure times. 
The \textit{Spitzer} photometry for our sources is given in Table~\ref{tabIR}.
We note that Src.~24 was not cross-identified with an unique IRAC 
counterpart because of the proximity of several infrared sources that contaminate 
the photometric measures. In the tables, we give photometry of the closest star to 
this Src.~24 for completeness. However, those values are not reliable and we did not 
used them in our study.

%_____________________________________________________________
\begin{table}
  \caption{Aperture correction factors for several different extraction radii.} 
  \label{tab2}    
  \centering       
  \tiny         
  \begin{tabular}{c c c c c}      
    \hline
    & \multicolumn{4}{c}{IRAC channel} \\
    \cline{2-5}
    Aperture & 3.6$\mu$m & 4.5$\mu$m & 5.4$\mu$m & 8.0$\mu$m \\
    (arcsec)  &     (mag)      &     (mag)      &     (mag)      &     (mag)      \\ 
    \hline      
    1.5 & 0.59 $\pm$ 0.04 & 0.63 $\pm$ 0.04 & 0.83 $\pm$ 0.04 & 0.94 $\pm$ 0.06 \\              
    2.0 & 0.32 $\pm$ 0.03 & 0.36 $\pm$ 0.03 & 0.53 $\pm$ 0.02 & 0.65 $\pm$ 0.03 \\
    3.0 & 0.13 $\pm$ 0.03 & 0.14 $\pm$ 0.02 & 0.22 $\pm$ 0.02 & 0.36 $\pm$ 0.02 \\
    4.0 & 0.03 $\pm$ 0.02 & 0.03 $\pm$ 0.02 & 0.05 $\pm$ 0.02 & 0.21 $\pm$ 0.03 \\
    \hline                                   
  \end{tabular}
\end{table}
%_____________________________________________________________

%_____________________________________________________________
 \begin{figure}
 \centering
 \includegraphics[width=8.0cm]{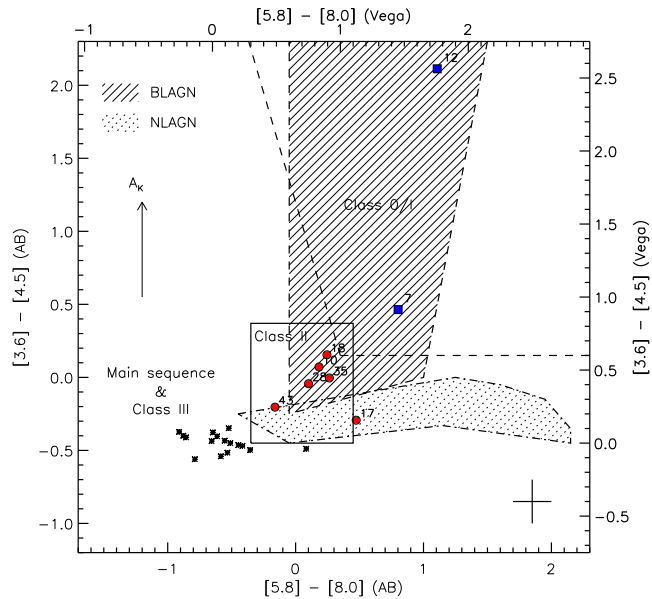}
   \caption{IRAC color-color diagram of the X-ray sources of NGC 2023 and its 
   surroundings. The locus of broad-line and narrow-line active galactic nuclei is 
   marked. Extinction vector is also overplotted. The cross symbol at the left bottom 
   corner indicates the typical error for sources in our sample. Squares are 
   sources classified as class I objects. Dots are class II objects while asterisks are 
   stars classified as main-sequence or class III objects. The extinction vector is $A_\mathrm{K} = 5$ mag.
   \label{irac_colors}}
 \end{figure}
%_____________________________________________________________

Figure~\ref{irac_colors} shows a color-color diagram of the four IRAC channels
for those of our sources with a counterpart in the four bands. We have marked the typical 
location of infrared class I, II and III objects and the locii of broad-line and narrow-line 
active galactic nuclei as shown by \citet{ste05}. Source ID numbers for class I and class II 
objects are overplotted. 
The cross at the right bottom corner represents the mean error bar in both colors. 
The typical error in the colors of bright sources is 0.1 mag. Faint sources show error bars 
with values as large as 1.2 mag in some cases.

The main concentration of sources is 
detected upon class III, that are usually identified with weak-line T Tauri and main sequence 
stars (asterisks in Figure~\ref{irac_colors}). Five sources are
situated inside the boundaries of the class II type objects and another two are very close to 
those boundaries. Stellar sources with infrared 
class II objects nature are typically identified with classical T Tauri stars. In Figure~\ref{irac_colors},
we have plotted them as filled circles. Finally, two sources have colors typical of class I 
objects (squares in the figure). Stellar sources with such infrared properties use to be 
related to protostars and T Tauri stars with circumstellar envelopes. A note of caution 
must be given here. Very absorbed class II objects may present IRAC colors typical of 
class I objects. Therefore, the two sources in Fig.~\ref{irac_colors} classified as class I 
objects may be absorbed class II stars. 
%, although they may be
%also classical T Tauri stars with highly absorbed by the presence of interstellar medium. 
%We used IRAC colors to classify stellar sources as class I, II or III objects 
%(see Fig.~\ref{irac_colors}). The location of the different classes is shown in the 
%figure. The locii of broad-line and narrow-line active galactic nuclei is also overplotted
%\citep[e.g.][]{ste05}. 
%We notice that Fig.~\ref{irac_colors} contains only IRAC counterparts of the X-ray 
%sources detected in the \textit{XMM-Newton} observation and not all the infrared 
%sources detected in the IRAC mosaics. 
%The diagram shows concentrations 
%mainly upon class III objects (weak-line T Tauri stars; asterisks in Fig.~\ref{irac_colors}) 
%or foreground main-sequence stars and  class II objects (classical T Tauri stars; dots in 
%the figure). Some sources with colors typical of class I objects (squares in the figure) are 
%also present. 
In fact, one of the sources has been previously classified as a class II star in the literature.
This is the case of the object detected by \citet{moo09} with Spitzer and SCUBA 
(named VLA~3 and MIR~46 in their paper, source number 7 in our work, Src~7). 
%The typical error in the colors of bright sources is 0.1 mag. 
%although 
%faint sources, mainly those with numbers 7, 13, 21, 31 and 46, show errors of up to 1.0 mag. 
%Source ID numbers for class I and class II objects are overplotted. 
%Note that the three sources with unclear classification (plotted as triangles in 
%Fig.~\ref{irac_colors}) are those with larger error bars. Their location in the color-color
%diagram is very uncertain.

%_____________________________________________________________
 \begin{figure}
 \centering
 \includegraphics[width=8.0cm]{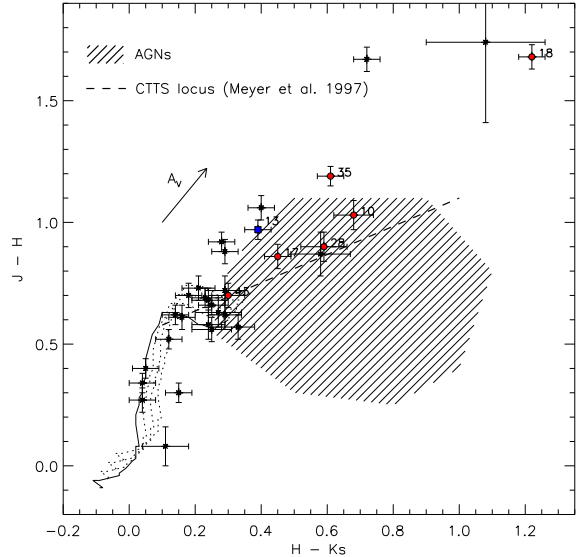}
   \caption{2MASS color-color diagram of the X-ray sources of NGC 2023 and its 
   surroundings. The dwarf stars main sequence is plotted as a continuous line. 
   The extinction vector is $A_\mathrm{V} = 3$ mag. The dotted lines correspond 
   to the dwarfs main sequence with $A_\mathrm{V} = 0.5$, 1.0 and 1.5~mag.
   Symbols are as in Fig.~\ref{irac_colors}.}
   \label{2mass}
 \end{figure}
%_____________________________________________________________

\subsection{Other photometric data from the literature}
\label{other_phot}

We complemented our study by cross-correlating our sample against different catalogs. 
In particular, we used the second version of the TASS Mark IV  photometric catalog \citep{dro06,droege2007}
for $V$ magnitudes, DENIS database 3rd release \citep{den05} for $I$ magnitudes, the 2MASS database 
\citep{skr06} for near-infrared, and the WISE preliminary data release \citep{cut11} for mid-infrared.
The cross-correlation was performed through the specific Aladin Java tool \citep{bon00}. We used 
a search radius of 5 arcsec to prevent the loss of coincidences due to the \textit{XMM-Newton} 
positional uncertainties.
We finally cross-matched our sample with the list of stars of \citet{moo09}.  

%There are 894 2MASS sources in the EPIC field of view, of which 144 have 
%uncertain photometry. Eighteen of our 50 X-ray sources present no counterpart in the 2MASS 
%database. 
Tables~\ref{tabNIR} and \ref{tabIR} summarize the photometric data for each one of our 
sources.  %In both tables, we mark spurious X-ray detections as `\textit{sp.}' Those are sources 
%detected in, or close to a detector gap in the \textit{XMM-Newton} EPIC and showing 
%no counterpart in any of the databases used for our study. There are 13 sources in our sample 
%identified with spurious detections. Another source 
We note that 13 out of the 50 sources detected in X-rays have no counterpart in 
any of the catalogs mentioned above. We considered them as spurious X-ray detections and, therefore, 
we removed them from the tables. We maintained Src. 23 because it
is situated at 11 arcsec from an optical 
source. We cannot discard this source as a real detection. The remaining sources have a 
counterpart in DENIS and/or 2MASS, except for source 7 \citep[VLA3 in][see Section~\ref{spitzer}]{moo09} and Src 12 (only \textit{Spitzer} photometry).
In Fig.~\ref{2mass} we plot the 2MASS color-color diagram for these sources. The main
sequence for unabsorbed (dwarf) stars is overplotted as a continuous line. 
The shaded region in the figure is the locus of the active galactic nuclei (AGN) 
determined by us using data from \citet{kou10}.
The position of most of the stars in the color-color diagram indicates the high interstellar extinction 
in this star-forming region. As in Fig.~\ref{irac_colors}, we identified class II objects with its X-ray 
source ID number. 
%
%From this figure we conclude that all the sources with 2MASS photometry are likely stellar objects. 
%Note that the locus of active galactic nuclei in the 2MASS color-color diagram is far at the right of 
%the star's main-sequence, while all our sources are over the main sequence or in the absorbed 
%main sequence path.
 %Note that some sources classified as class I or II objects in Fig.~\ref{irac_colors} are inside the 
 %boundaries of the AGN locus. This is not surprising since T Tauri stars and AGN show very similar 
 %characteristics in the infrared.}
 
%Table~\ref{tabIR} summarizes the infrared characteristics of our stellar sample. 
%Some X-ray sources have no infrared counterpart in one or both of them.
%They are very close to an \textit{XMM-Newton} detector gap or a bright X-ray source.
%They are very probably spurious sources and  are marked with an asterisk in the table. Note 
%that for some of them, \textsc{daophot} gives a value for their flux density in some IRAC channels. 
%IRAC flux density and 2MASS photometry are given for the remaining sources together with
%their classification as Class I, II and III objects extracted from Fig.~\ref{irac_colors}.  

We estimated the value of the extinction ($A_\mathrm{V}$) for each star, from the 2MASS 
color-color diagram (Fig.~\ref{2mass}). Our results are given in Table~\ref{tabNIR}. They will be discussed
and compared with those obtained from SED fitting in Section~\ref{SEDs}.

\begin{figure}
 \centering
 \includegraphics[width=8.0cm]{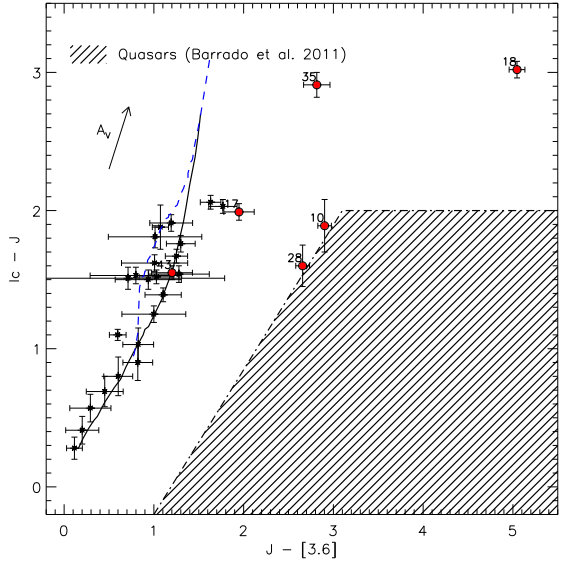}
   \caption{$I - J$ - $J - $[3.6] color-color diagram of the X-ray sources of NGC 2023. 
   The continuous line represents the ZAMS and the dashed line is the 1 Myr isochrone
   \citep{bar98}. The shaded region is the locus of quasars
   \citep[adapted from][]{bar11}. Symbols are as in Fig.~\ref{irac_colors}. The extinction 
   vector is $A_\mathrm{V} = 1.5$ mag.}
   \label{IJ_JL}
 \end{figure}

\citet{bar11} showed that the $I - J$ vs $J - $[3.6] color-color diagram can be used to reject
extragalactic objects from X-ray selected samples, as active galactic nuclei are well separated 
from stars. Figure~\ref{IJ_JL} shows the same diagram for our 32 sources with a DENIS 
counterpart. Src~10 and Src~28 are situated close to the boundaries of extragalactic sources, 
but they have been clearly identified with stars in the optical images (in fact Src~10 is V615~Ori). 
Only Src~18 and Src~35 may be absorbed quasars, according to the direction of the extinction vector
(In Tables~\ref{tabNIR} and \ref{tabIR} they are identified as possible AGN).

%Finally, we estimated the value of the extinction ($A_\mathrm{V}$) for each star, from the 2MASS 
%color-color diagram (Fig.~\ref{2mass}). Our results are given in Table~\ref{tabNIR}. They will be discussed
%and compared with those obtained from SED fitting in Section~\ref{SEDs}.

%_____________________________________________________________
%_____________________________________________________________

\subsection{Age of the NGC 2023 members}
\label{age}

It was mentioned in Section~\ref{intro} that the age of the star-forming region around 
NGC 2023 is not well known. There are very few works about this issue in the literature 
\citep[see][for a review]{mey08}. \citet{alc00} gave a range of ages for the stars belonging to 
the Orion Molecular Complex, but there is no particular study on each individual region. 
Ages determined for these stars are in the range 0.5--7.0 Myr. %what makes comparing their
%properties with those of members of other star-forming region difficult.

%_____________________________________________________________
 \begin{figure}
 \centering
 \includegraphics[width=8.0cm]{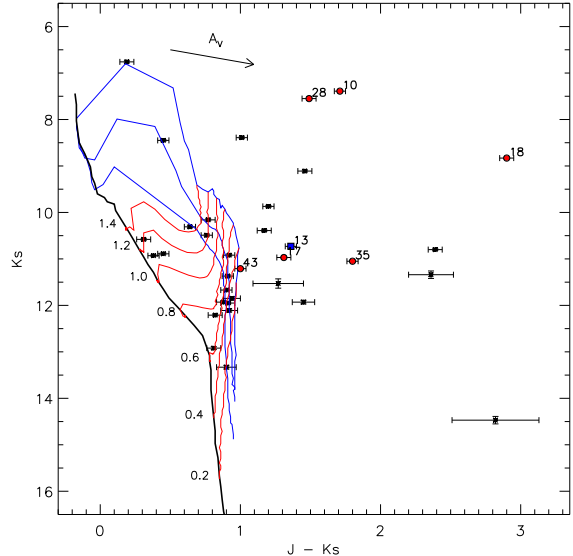}
   \caption{$K_{s}$ - $J - K_{s}$ color-magnitud diagram of the X-ray 
   sources of NGC 2023. The lines are evolutionary tracks from \citet{bar98} and \citet{sie00}, 
   with the ZAMS, 7, 3 and 1 Myr isochrones and several mass tracks ($d = 400$ pc). 
   Symbols are as in Fig.~\ref{irac_colors}. The extinction vector is $A_\mathrm{V} = 5$ mag.}
   \label{K_JK}
 \end{figure}
%_____________________________________________________________

In Fig.~\ref{K_JK}, we show evolutionary tracks from \citet{bar98}, with an extension to stellar 
masses over 1.6 M$_\odot$ from \citet{sie00}, in a 2MASS color-magnitud diagram.
We assumed a mean distance of 400 pc for the region. The isochrones are 1, 3 and 7 
Myr and the ZAMS. We also plot tracks for masses ranging from 0.2 to 1.4 M$_\odot$. 
%Many stars seem to follow the $7$ Myr isochrone, while other stars seem to be 
%younger if they are assumed to be at that distance, although this may be an artifact produced
%by a combination of both an incorrect distance and a higher extinction.
Although there is a concentration of low mass stars around the 7~Myr isochrone, the density of 
isochrones in that region of the diagram makes difficult to assure this is the age for the cluster.
In fact, the most massive stars seem to follow younger isochrones. Therefore, we only 
may affirm the age of the stars in NGC~2023 is in the approximate range $1-7$~Myr.

\subsection{SED fitting}
\label{SEDs}

To investigate in more detail the nature of our X-ray sources, we performed SED fitting using
the photometric data we compiled for each source (Tables~\ref{tabNIR} and \ref{tabIR}). We took 
full advantage of the grid of young stellar object radiation transfer models of \citet{rob07}. In the model, 
we permitted the extinction to vary only from zero to the maximum value given by the 2MASS color-color 
diagram (Fig.~\ref{2mass}). Extinction and distance are the only parameters external to the model. 
%The remainder (stellar mass and luminosity, protostellar disk and circumstellar envelope parameters) 
%is part of the radiation transfer model.  

Fitting was performed for the 32 sources in our sample with a DENIS counterpart, to include 
at least one optical band. This should improve spectral type determination. The blue 
spectrum of a %classical T Tauri star 
Class II is usually affected by accretion, as it produces an excess in the blue 
and ultraviolet spectral range. %Although the \citet{rob07} model includes a treatment of this excess, the 
%lack of ultraviolet data for our sources may produce systematic errors in the determination of spectral types.  
%%We have studied also the SED of Src.~24 that lacks a DENIS counterpart but has counterparts in 2MASS 
%%and \textit{Spitzer}.
%
In Figs.~\ref{seds_classIII} and \ref{seds_classII}, we show the best fit for those 32 sources. 
In some cases, there is more than one model that fits well our data. For those cases, we  
overplot them as grey lines. We have assumed as good models those in which  
($\chi^2 - \chi^2_\mathrm{best})/\chi^2_\mathrm{best} \le 0.1$. 
In most of those cases, it is not possible to discard any of the 
fits. We would need some point in the sub-millimeter spectral range (see discussion in Section~\ref{discussion}). 

The classification into the different infrared classes given in Tables~\ref{tabNIR} and ~\ref{tabIR}
was performed according to the results of the SED fitting, using the parameter $\alpha$, defined as the 
slope of the SED beyond 2.2~$\mu$m \citep[see][]{rob07}. We used every point of the SED for this 
purpose, joining WISE and \textit{Spitzer} data.  A review on the classification method can be found in 
\citet{sta05}.
We classified as class III objects those sources for which an absorbed stellar model is enough 
to fit the observed SED ($\alpha \le -1.5$). Only Src.~14 may show a very thin circumstellar disk 
(see Fig.~\ref{seds_classIII}). Nevertheless, the data fits well also to a stellar photosphere.
%and with 
%the lack of points at longer wavelength we cannot assure this star has a disk.  
%
Class II objects ($-1.5 < \alpha \le 0.0$) show an obvious excess at the IRAC bands (see Fig.~\ref{seds_classII}). 
For those sources, any attempt of fitting a stellar model without disk was unsuccessful. 
Src.~13 and Src.~17 show characteristics of both class I and II objects. Their SED first decreases between 1.0 and 
5.4 $\mu$m, but then increases again. We have classified them as class I/II objects. %(see the discussion 
%section). 
%
Finally, there are several sources that have SEDs typical of class III objects but 
with an excess at wavelengths longer than 10~$\mu$m. We have classified them as class II/III objects. 
Src~42 is situated in the IRAC color-color diagram 
(Fig.~\ref{irac_colors}) in a position that is usually occupied by stars with transition disks.
Our classification into infrared classes using the SED coincides very well with that made first 
using the IRAC color-color diagram, except for Src~44, the SED of which shows an excess above 10~$\mu$m 
but that in the color-color diagram is situated inside the boundaries of class III objects. 
%Our study of the source SEDs is undoubtedly more complete than that performed using only IRAC colors 
%since it contains all the detected X-ray sources except for the two of them with no counterparts in 
%2MASS (Src~7 and Src~12) and Src~16, for which we have only photometry of 2MASS.  

Finally, we note that the SED fitting tool of \citet{rob07} gives an age for each star. This value is determined 
through isochrone fitting, so its result depends strongly on the model chosen. \citet{rob07} use the \citet{sie00} 
pre-main sequence evolutionary models that differ slightly from the \citet{bar98} models. The tool also 
tends to fit low-age isochrones even when a simple photosphere model is suitable for the observed SED.
Therefore, the results of the fitting tool  should be used with some caution.

Ages determined using the \citet{rob07} fitting tool are in the range 5--9 Myr for most 
of the stars, what may agree with the results presented in Section~\ref{age}. Nevertheless, for stars 
showing larger infrared excesses, ages are in teh range 0.5--2 Myr.

\subsection{Extinction, stellar masses, temperatures and luminosities}

Table~\ref{tabSEDs} summarizes the results of the SED fitting for the 32 sources with 
an optical counterpart (see Section~\ref{SEDs}). For sources with more than one 
good fit, %(considering good fits those for which ($\chi^2 - \chi^2_\mathrm{best})/\chi^2_\mathrm{best} \le 0.1$), 
we give only parameters for the best fit. For class II objects, the accretion rate and evolutionary stage 
are also given. To account for the evolutionary stage of young stellar objects,
\citet{rob07} defined a parameter based on physical properties of the 
circumstellar envelops, such as mass and accretion rate from it. The authors distinguished 
between stars with significant infalling envelops (Stage 0/I objects), stars with optically thick disks 
and thin envelopes
(Stage II objects), and stars with optically thin disks (Stage III objects).  We have determined this stage
for our sources and included it in Table~\ref{tabSEDs}. We observe that this classification coincides very 
well with that given by the IRAC color-color diagram or the slope of the SED. Class I objects present 
properties of Stage I objects while class II objects are classified as Stage III objects following this method.

We previously mentioned that parameters listed in Table~\ref{tabSEDs} are those given by the model. 
In fact, the extinction (A$_\mathrm{V}$) and the distance are parameters external to the radiative model 
\citep[see][]{rob07}. In addition, stellar mass and radius are determined from evolutionary tracks 
\citep{rob08}. In essence, the parameters derived from the model are the temperature and luminosity
of the star. %Therefore, the remaining \textbf{values} given in Table~\ref{tabSEDs} should be 
%used with caution. %For sources for which a simple stellar photosphere fits the SED correctly, the 
%fitting tool do not derive masses luminosities and radii. Therefore, we used the zero-age main-sequence 
%track of \citet{sie00} to determine them.

With respect to the interstellar extinction, the values coming from the SED fitting tool are quite similar to those 
determined with the 2MASS color-color diagram if one takes error bars into account for stars showing low extinction. 
For the remaining stars, there is a trend of $A_\mathrm{V}$ determined using the color-color diagram to be higher 
than that coming from the SED fitting tool. The median of the ratio between $A_\mathrm{V}$ determined with both 
methods is 1.35, or 1.5 if only stars with $A_\mathrm{V} \ge 1$ are considered. These stars are mainly those for 
which the extinction determined with both methods do not coincide even when error bars are taken into account.

%_____________________________________________________________________________________________________________________
\begin{table*}
\caption{Results of the X-ray spectral fitting to hot plasma models for stars in NGC 2023. Errors are at 90\% 
confidence level.}    
\label{xspec}
\centering                
\begin{tabular}{l c c c c c c c c}      
\hline              
    ID &  $N_\mathrm{H}$ &  $kT_1$ & $kT_2$ & $EM_1/EM_2$ & Z & $\chi^2$ & 
    $f_\mathrm{X}$$^{a}$ & $\log L_\mathrm{X}$$^{b}$\\
   &$(\times10^{22}$ cm$^{-2})$ & (keV) & (keV) &  & (Z$_{\odot}$) & (d.o.f) & $(\times10^{-13}$ erg cm$^{-2}$ s$^{-1}$) & (erg s$^{-1}$)\\
\hline 
\noalign{\smallskip}
3 & 0.00$_{-0.00}^{+0.01}$ & 0.62$_{-0.11}^{+0.09}$ & 1.26$_{-0.13}^{+0.21}$ & 0.70 & 0.23$_{-0.05}^{+0.05}$ & 1.29 (276) & 3.95 & 29.2 \\
\noalign{\smallskip}
10 & 0.50$_{-0.04}^{+0.05}$ & 2.49$_{-0.52}^{+0.34}$ & ... & ... & 0.33$_{-0.16}^{+0.22}$ & 1.29 (140) & 12.20 & 31.3 \\
\noalign{\smallskip}
17 & 0.22$_{-0.08}^{+0.44}$ & 1.24$_{-0.53}^{+1.62}$ & ... & ... & $=0.2$ & 1.27 (54)& 0.57 & 30.3\\
\noalign{\smallskip}
20 & 0.07$_{-0.05}^{+0.05}$ & 0.78$_{-0.80}^{+0.16}$ & 1.64$_{-2.32}^{+2.32}$ & 2.59 & 0.13$_{-0.03}^{+0.13}$ & 0.88 (78) & 2.30 & 30.5 \\
\noalign{\smallskip}
22 & 0.05$_{-0.04}^{+0.06}$ & 0.34$_{-0.04}^{+0.07}$ & 1.21$_{-0.20}^{+0.17}$ & 1.29 & 0.18$_{-0.06}^{+0.10}$ & 0.83 (120) & 3.30 & 30.5\\
\noalign{\smallskip}
27 & 0.41$_{-0.21}^{+0.33}$ & 18.3$_{-13.7}^{+60.0}$ & ... & ... & $=0.2$ & 0.51 (10) & 1.38 & 30.0\\
\noalign{\smallskip}
28 & 0.44$_{-0.05}^{+0.07}$ & 0.28$_{-0.04}^{+0.05}$ & 1.24$_{-0.05}^{+0.04}$ & 1.02 & 0.16$_{-0.03}^{+0.03}$ & 1.10 (602) & 33.41 & 31.9 \\
\noalign{\smallskip}
34 & 0.40$_{-0.08}^{+0.10}$ & 2.73$_{-0.58}^{+1.08}$ & ... & ... & $=0.2$ & 1.26 (59) & 6.99 & 30.9\\
\noalign{\smallskip}
36 & 0.24$_{-0.10}^{+0.16}$ & 2.14$_{-0.81}^{+1.76}$ & ... & ... & $=0.2$ & 0.72 (61) & 2.17 & 30.4\\
\noalign{\smallskip}
41 & 0.08$_{-0.02}^{+0.03}$ & 2.36$_{-0.61}^{+0.55}$ & ... & ... & 0.15$_{-0.10}^{+0.14}$ & 1.18 (271) & 11.53 & 30.6 \\
\noalign{\smallskip}
50 & 0.22$_{-0.08}^{+0.08}$ & 2.00$_{-0.62}^{+1.86}$ & ... & ... & $=0.2$ & 1.33 (65)& 3.41 & 30.2\\
\noalign{\smallskip}
\hline
\end{tabular}
\flushleft{$^{(a)}$Unabsorbed flux in the energy band [0.3-8.0] keV. $^{(b)}$ Unabsorbed X-ray luminosity in the energy band [0.3-8.0] keV. 
We assumed $d = 450$ pc (the distance to the ONC) for all the sources. }
\end{table*}
%_____________________________________________________________________________________________________________________

\section{X-ray spectral analysis} 
\label{fit}

%In Section~\ref{obs}, we showed that the X-ray observation of the NGC 2023 nebula was affected 
%by high background periods. The high-energy lightcurve of the EPIC-PN detector is shown in 
%Fig.~\ref{fig2}. Shaded regions in the plot are periods of highly variable X-ray background. 
%For our spectral analysis, we discarded these periods. With this procedure, we assured that the 
%individual source spectra were not contaminated with high-energy photons. In addition, the highly
%variable background prevented us to perform a study about variability of the detected sources. 
%Therefore, our study was limited to the spectral analysis. 

In Section~\ref{obs}, we showed that the X-ray observation of the NGC 2023 nebula was affected
by high background periods (see Fig.~\ref{fig2}). The highly variable background prevented 
us to perform a study about variability of the detected sources. Therefore, our study was limited to 
the spectral analysis.
The X-ray spectrum of  sources with more than approximately 100 net counts was compared (using 
fitting procedures) to  
hot plasma models using XSPEC \citep{arn96,arn04}.  The software first load the 
data previously extracted using specific SAS tools (source and background spectra, 
response matrix and ancillary file) and then generates the plasma model. Two hot (diffuse) 
plasma models are available in XSPEC: \textit{mekal} \citep{mew85,mew86,kaa92,lie95} 
and \textit{apec} \citep{smi01}. To be consistent with our previous works \citep[e.g.][]{lop08,lop10a},
we chose the latter for the present analysis. The {\it apec} uses the atomic data contained in 
the Astrophysical Plasma Emission Database \citep[APED;][]{smi01b}. %, which is being continuously 

Due to the limitations of our observation in terms of counts --\,because of the  
time lost due to the high background periods\,-- we used only $1T$ or $2T$-models for the fits. 
%This is a typical approach for low-resolution X-ray spectra and permits to compare results 
%with the close star-forming region NGC~2024 \citep{ski03}.
%
The results of the spectral fitting are given in Table~\ref{xspec} for those stars with accurate fits. 
For several stars, the value of the abundance obtained during the fit was unconstrained. In those 
cases, we fixed the abundance to $Z/Z_{\odot} = 0.2$. Although some authors have used other values
in their works \citep[e.g.][]{ste04,robrade07,gia07}, in other stars of this star-forming region we 
find always a value of $Z/Z_\odot$ very close to 0.2. 
We used $\chi^2$ Statistics to obtain the goodness of our fits. Energy channels were then 
grouped to contain at least 15 counts in each energy bin. 

%Although many authors choose this 
%number for their fits, it is not mandatory and, in fact, it is not recommended since spectral information 
%is lost (K. Arnaud, in private communication). In fact, the $\chi^2$ test should be applied when 
%the distribution of points in each bin is assumed to be a sum of Gaussian distributions.
%%(see Section~\ref{chi2}). 
%Deviations of the observed counts from the expected ones (where 
%the expected values are those coming from the model) would be due to statistical fluctuations only. 
%%Finally, for our work we decided to use the $\chi^2$ test in XSPEC with a grouping of 15 counts/bin 
%%to be consistent with other works in the literature. 

\section{Discussion}
\label{discussion}

Spectral fitting was performed for the eleven sources in our sample with at least 500 counts (combining PN and 
MOS detectors). The results are shown in Table~\ref{xspec}. %A quick comparison of our results for class III stars (see 
%Table~\ref{tabNIR}) with those coming from comparing the measured hardness ratio for the same sources with models 
%(Fig.~\ref{hr}), shows that they are compatible. Instead, for %classical T Tauri stars 
%Class II and other stars with high Hydrogen 
%column densities, the direct comparison of measured harness ratio with models underestimates the plasma temperature 
%and overestimates $N_\mathrm{H}$ (see, for instance, results for Src~10 and Src~17). 
%This result indicates that 1T-models are not able to reproduce the X-ray spectrum of such stars.

The position in the color-color diagram of Fig.~\ref{IJ_JL} of the eleven spectroscopically sources studied in this work  
indicates they are indeed stars. According to that diagram, only Src~18 and Src~35 may be (absorbed) extragalactic 
sources. 
General X-ray parameters obtained for these eleven stars are very similar to those observed in stars of the same class 
in other star-forming regions \citep[e.g.][]{getman05,gudel2007}. Among them, three are classified as 
%classical T Tauri stars 
Class II from their SEDs (see Tables~\ref{tabNIR} and \ref{tabIR}.). They are Src~10 (V615~Ori), 
Src~17 and Src~28. Another two stars likely have transition disks (Src~36 and Src~41). 
%High column densities are observed in the analyzed class II objects (sources 10, 17, and 28) 
%although three sources classified as 
The remainder are classified as %weak-line T Tauri 
 Class III or main-sequence stars. %Nevertheless, Src~27, classified as a 

Coronal temperatures are consistent with being young T Tauri stars \citep[e.g.][]{gudel2007}. Only Src~27 
shows an extremely high temperature, although it is unconstrained (this is the source with the 
lowest number of counts we could fit). 
Unabsorbed X-ray fluxes in the energy band [0.3--8.0] keV were determined for the eleven sources 
for which a spectral fitting was performed and then transformed to luminosities 
%assuming a mean distance of 450 pc, as for the Orion Nebula Cluster (ONC). 
using the distances obtained during the SED fitting procedure (Table~\ref{tabSEDs}).
We note that for HD~37805 (Src~3), the distance given by the SED fitting tool is slightly underestimated 
(Hipparcos distance for HD~37805 is $d = 82 \pm 7$ pc). However, this is the only star for which we could 
find a value for its distance in the literature. Therefore, an interpolation of this result to the remaining stars should 
not be done. In fact, the distance obtained with this procedure for V615~Ori (Src~10, $d = 371$ pc) is consistent 
with some results found in the literature: there are hints that the Horsehead Nebula is a few parsecs closer to us than 
$\sigma$~Orionis, that is situated at approximately 385 pc \citep[see][and references therein]{cab08}. 
V615~Ori must be closer than the Horsehead Nebula as it is observed at the base of the nebula, 
in front of it. Therefore, the use of the distances obtained from the SED fitting tool seems to be a good option, 
undoubtedly better than using the same distance for all the sources. Observed X-ray 
luminosities are indicative of the shallowness of this observation. %, incorrect 
%distance for the sources or even a spread in age of these regions.
Typical X-ray luminosities of young M dwarfs are in the range $L_\mathrm{X} \sim 10^{28.5} - 10^{30}$ erg\,s$^{-1}$
%\citep{lop10a}
\citep[e.g.][]{pre05a}. In this observation, the completeness limit is $f_\mathrm{X} ~\sim 5 \times 10^{-14}$ erg\,cm$^{-2}$\,s$^{-1}$, 
what corresponds to $L_\mathrm{X} \sim 10^{30}$ erg\,s$^{-1}$ at a distance of 400 pc. 
%Nevertheless, these values 
%are similar to those found in other T Tauri stars  \citep[e.g.][]{gudel2007,arg09}.
\citet{ste12} pointed out  that there is a trend of a mass dependence of the X-ray emission level throughout the various 
evolutionary stages of young stellar objects. Comparing the luminosities of class II and class III objects in our sample, we do 
not find such differences, but our sample is too small to achieve robust results.
 \begin{figure}
 \centering
 \includegraphics[width=8.5cm]{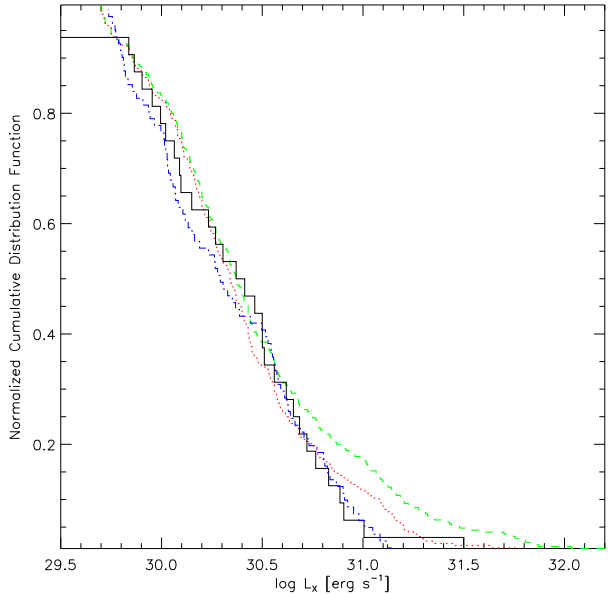}
   \caption{\bf Cumulative distribution function of  $\log L_\mathrm{X}$ in NGC 2023 (continuous line, 
   black), Taurus (dashed-dotted line, blue) and in the Orion Nebula Cluster (ONC). 
   The ONC is plotted twice, one for the whole sample (dashed line, green) and 
   the other with stars with masses less than 3 M$_\odot$, plotted as a dotted (red) line.}
   \label{XLF}
 \end{figure}
%_____________________________________________________________

With the aim of comparing general results of our X-ray analysis of NGC 2023 with other star-forming regions, 
we plotted the X-ray luminosity cumulative distribution function of our sample, the Orion Nebula Cluster 
and the Taurus Molecular Complex together (see Fig.~\ref{XLF}). The data of those regions were taken 
from \citet{getman05} and \citet{gudel2007}, respectively. For each star-forming region, we 
plot the whole sample accessible in those works with no distinction between infrared classes.
%For instance, 
%\citet{ski03} shows in their Fig.~15, the luminosity cumulative distribution of some class II objects. However, we
%use all the sources in their Table~1.
%
For NGC 2023, we discarded 
HD~37805 (Src~3), that is likely a field star, and those stars for which we did not fitted a model to their 
SEDs (Src~7, Src~12, Src~16 and Src~24).  %Those sources for which 
%no infrared counterpart was found (marked with * in Table~\ref{tabIR}) were also discarded. 
For the stars for which a spectral analysis with XSPEC could not be performed, we determined X-ray fluxes from 
the observed count-rate in the following way. We used the subsample of eleven stars in Table~\ref{xspec} for which X-ray 
spectral fitting was done to determine a conversion factor between observed count-rate and (absorbed) flux. This 
conversion factor was determined for each star and each EPIC detector (PN, MOS1 and MOS2) independently. Then, 
we performed a linear regression to each one of the three datasets. %For MOS1 and MOS2, a simple linear regression 
%fits the subsample data, but for the PN data, a cubic polynomial was needed (see Fig.~\ref{cf}). 
The relationships obtained by us were:

\begin{eqnarray*}
f_X[MOS1] & = & (0.5 \pm 0.3) + (0.056 \pm 0.005)  \cdot CR \\
f_X[MOS2] & = & (0.2 \pm 0.4) + (0.057 \pm 0.005)  \cdot CR \\
f_X[PN] & = & (1.5 \pm 0.2) + (0.003 \pm 0.005)  \cdot CR %- \\
              %&    & (0.5 \pm 2.2) \cdot 10^{-5}  \cdot CR^2 + \\
              %&    & (4.0 \pm 2.9) \cdot 10^{-8}  \cdot CR^3
\end{eqnarray*}

%From Fig.~\ref{cf}, we notice that stars with positive values of HR$_1$ (see Section~\ref{obs}), seem to follow a different 
%trend. Nevertheless, the three sources with positive HR$_1$ values in the whole sample are Src~10 (V615~Ori), Src~27 
%and Src~34 which spectrum was analyzed with XSPEC. Therefore, we fitted the different polynomials only to the remaining 
%eight stars of Table~\ref{xspec}.
%
%Finally, we checked that
Fluxes determined with these three relationships are compatible. %Hence, we applied them to 
%our sample of 36 sources to determine their flux. 
Table~\ref{tabX} gives the flux values determined for our sample from their count-rates. We include in that table the 
eleven stars of Table~\ref{xspec}, for completeness. For them, the values of the fluxes are those determined from 
the spectral fitting. As for the eleven stars for which we could perform spectral fitting (see above), fluxes 
were transformed into luminosities using the distances obtained from the SED fitting (Table~\ref{tabSEDs}). 
%A correction factor of 1.78 was applied to the luminosities that accounts for the mean absorption found in this region.

%_____________________________________________________________
 %\begin{figure}
 %\centering
 %\includegraphics[width=8.5cm]{cf.ps}
 %  \caption{Count-rate to flux conversion factor relationships determined for the stars
 %  in Table~\ref{xspec} using EPIC PN count-rates. Dotted line is the linear regression, 
 %  dashed line is a two-degrees polynomial and the continuous line is the three-degrees 
 %  (cubic) polynomial. The filled circles are stars with negative values of the hardness ratio HR$_1$ 
 %  in Table~\ref{tabX}, while open circles have positive values of HR$_1$ (see definition of 
 %  HR in Section~\ref{obs}).}
 %  \label{cf}
 %\end{figure}
%_____________________________________________________________

The comparison of the X-ray luminosity function of the different star-forming regions 
was made only in the range of completeness of the \textit{XMM-Newton} observation of NGC 2023
($\log L_\mathrm{X} [\mathrm{erg\,s}^{-1}] \ge 29.8$).
Known non-members of the ONC identified by \citet{getman05} were discarded for this study.
From Fig.~\ref{XLF}, NGC~2023 seems to be more similar to the Taurus molecular complex. In both cases, 
there is a lack of stars brighter than $L_\mathrm{X} \sim 10^{31}$ erg\,s$^{-1}$, while %NGC~2024 and 
the ONC have stars with X-ray luminosities up to 10$^{31.5-32}$ erg\,s$^{-1}$. This result may be explained 
as differences in the stellar population in each star-forming region. While the ONC have massive stars,
they are not present in Taurus and NGC~2023. We have explored this hypothesis by constructing a new 
COUP sample with stars less massive than 3~M$_\odot$ (similar to the mass range in NGC~2023). 
In Fig.~\ref{XLF}, this new sample is plotted as a dotted (red) line. The figure suggests that the COUP 
sample restricted to the same mass range is more similar to NGC 2023 than the sample without mass
restriction. We note here that the ages of the Taurus and ONC samples are not equal and that 
we could not determine robustly the age of NGC~2023. Therefore, this result must be used with caution.

\begin{figure}
 \centering
 \includegraphics[width=8.5cm]{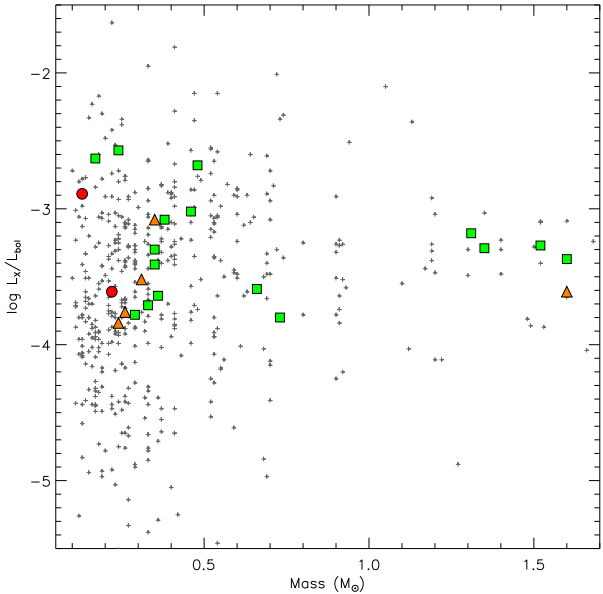}
   \caption{$\log L_\mathrm{X}/L_\mathrm{bol}$ as a function of stellar mass for the stars of the COUP 
   catalog (small crosses) and our sample of NGC~2023 stars. Squares are stars classified as class III 
   in Tables~\ref{tabNIR} and ~\ref{tabIR}. Filled circles are infrared class II objects. Triangles are class II/III
   objects.}
   \label{lxlbol}
 \end{figure}

In Fig.~\ref{lxlbol}, we overplot our sample of NGC~2023 sources over the COUP data \citep{getman05}. 
For NGC~2023, we used the masses and bolometric luminosities given in Table~\ref{tabSEDs} and X-ray luminosities 
determined from the distances given in Table~\ref{tabSEDs} and fluxes of Table~\ref{tabX}. All the NGC~2023
low-mass stars ($M < 1 M_\odot$) are inside the boundaries of the stars of the COUP.  Only the more massive stars 
in our sample \citep[as determined by the SED fitting tool of][]{rob07} have very low values of 
$\log L_\mathrm{X}/L_\mathrm{bol}$ (out of the limits of Fig.~\ref{lxlbol}, see Table~\ref{tabSEDs}). 
There are three types of sources with very low values of $\log L_\mathrm{X}/L_\mathrm{bol}$ in our sample: 
(1) stars showing large infrared excess, for which the SED fitting tool gives very high luminosities, 
(2) intermediate-mass pre-main or main-sequence stars (A and F spectral types), and (3) Src.~18, that may be 
an AGN instead of a star (see Section~\ref{other_phot} and Fig.~\ref{IJ_JL}). The SED fitting procedure 
derives bolometric luminosities and temperatures directly and then determine masses using evolutionary 
tracks \citep[see][]{rob08}. For stars with large infrared excesses, to obtain a robust value for their luminosity 
may be difficult since it depends on the accurate determination of the disk (and envelope in some cases) 
parameters. Therefore, in those stars of our sample, $L_\mathrm{bol}$ may be overestimated. 

For the two stars classified as infrared class III objects (i.e. stars without an accretion disk) 
their temperature may be also overestimated. Nevertheless, the star HD~37805 (Src.~3) that is one of these 
two pre-main or main-sequence stars with a very low value of the ratio $\log L_\mathrm{X}/L_\mathrm{bol}$, 
it is a known A star \citep[e.g.][]{moo09}, already detected in X-rays with ASCA \citep{yam00}. Therefore, the 
presence of a low-mass companion responsible for the X-ray emission cannot be discarded in these three
intermediate-mass stars in our sample. The other class~III object with low $\log L_\mathrm{X}/L_\mathrm{bol}$
is Src~33 (MIR~83).

%_____________________________________________________________
%_____________________________________________________________

%In general, there is a good agreement between the cumulative distribution functions of the ONC, Taurus, 
%NGC 2024 and NGC 2023, for $\log L_\mathrm{X} [\mathrm{erg\,s}^{-1}] \le 30.5$.  However, our sources 
%seem to be over-luminous  
%above that value. While for the ONC, Taurus and NGC 2024 the cumulative distribution function 
%decreases softly towards high X-ray luminosities, for NGC 2023 it changes abruptly, showing 
%more high luminous sources than expected from the results of the other star-forming regions. As we 
%discarded known and possible field stars where the wrong use a of distance of 450 pc had would lead to the 
%overestimation of their X-ray luminosity, we suggest that this result is not produced by the use of a wrong 
%distance. A possible explanation for the differences in the cumulative distribution function of NGC 2023 is  
%the occurrence of large flares during the observations. We notice that we could not carried out a 
%variability study due to the high variability of the X-ray background during the \textit{XMM-Newton}
%observation. The second hypothesis is that those sources are indeed intrinsically more luminous. 

\section{Summary and conclussions}

We have carried out a comprehensive analysis of the X-ray and infrared properties of 36 young stellar objects 
in the NGC 2023 star-forming region and its surroundings for the first time. The spectral energy distribution 
(SED) of 90\% of this sample (32 out of 36 sources) has been analyzed using the star-disk models of \citet{rob07}. 
We have identified only two AGN candidate among the 36 X-ray sources. A classification on different 
infrared classes has been made according to the position of the stars in infrared color-color diagrams 
and to their SED. The study of the SEDs has allowed us to determine several stellar parameters such as 
masses, radii and luminosities, and other physical parameters such as extinctions and distances. 

The X-ray properties of NGC~2023 have been compared with 
those of other star-forming regions with different characteristics in terms of stellar population.
Although the values of temperature and column density determined for our sample of young stars are similar 
to those observed in stars of the Orion Nebula Cluster and Taurus, the (X-ray luminosity) cumulative distribution 
function of members of NGC~2023 is more similar to that of the Taurus molecular complex. 
A possible explanation is the different stellar populations in each region. While in the ONC
there is a number of massive stars, they are not present in Taurus and NGC~2023.

%significantly different from the other star-forming regions. In particular, NGC 2023 show 
%a bump in the CDF at high luminosities. We detected three class I objects and two more candidates. 
%Compared to the class 0/I objects detected in Orion and $\rho$ Ophiuchi, our class I objects seem to be 
%under-luminous in X-rays. The star-forming region surrounding NGC 2023 presents environmental 
%conditions that are different from those observed in other star-forming regions. In particular, it has 
%not OB stellar systems that could ionize the surrounding medium but it presents a considerable 
%number of protostars with outflows. Those differences may condition the presence of a larger number
%of high X-ray luminous sources in the region, for instance a large number of intermediate-mass 
%stars. Nevertheless, this hypothesis must be still contrasted with a deeper investigation of this 
%star-forming region. 

\section*{Acknowledgments}
 M.A.L-G. and J.L-S. acknowledge support by the Spanish Ministerio de Ciencia 
 e Innovaci\'on under grant AYA2008-06423-C03-03. J.F.A.C. is researcher of 
 the CONICET and acknowledges support by grant PICT 2007-02177 (SecyT).
  PGP-G acknowledges support from the Spanish Programa Nacional de Astronom\'{\i}a 
 y Astrof\'{\i}sica under grants AYA2009-10368 and AYA2009-07723-E.
 We would like to acknowledge the anonymous referee for a fruitful discussion about 
 the results of this work. His/Her comments on the text content have allowed us
 to notoriously improve this manuscript.

\appendix

\section{Tables and Figures}
\label{appendix}

\begin{table*}
\caption{X-ray parameters from the SAS multitask \textit{edetect\_chain}.}
\label{tabX}
\small
\centering      
  \begin{tabular}{l c c r r r r c c l}      
    \hline              
\noalign{\smallskip}
Src ID & $\alpha$  & $\delta$  & rate [pn] & rate [mos1] & rate [mos2] & Obs. flux$^{a}$ & HR1 & HR2 & Remarks$^{b}$\\
            & (deg)   & (deg)  &  (ks$^{-1}$) & (ks$^{-1}$) & (ks$^{-1}$) & ($\times 10^{-13}$ erg\,cm$^{-2}$\,s$^{-1}$) \\ 
\hline
\noalign{\smallskip}
1 &  05 40 54.2 & -02 19  04.8 &  ...	 &  8	$\pm$ 1 &  10	$\pm$ 2 &  0.79 $\pm$  0.44 & -0.55 &  -0.49 \\
%%2 &  05 41  02.4  & -02 18 28.8 &  ...	 &  7	$\pm$ 2 &  ...	 &  0.89 $\pm$  0.39 &  -0.14	 &  -0.83 & sp.  \\
%2 &  05 41  02.4  & -02 18 28.8 & ... & ... & ... & ... & ... & ... & sp.  \\
3 &  05 41  02.6 & -02 18 18.0 &  310 $\pm$ 8 &  67	$\pm$ 5 &  87 $\pm$ 4 &  3.95 $\pm$ 0.10	 &  -0.72 &  -0.80 & [YKK2000] A5 \\
4 &  05 41  03.8 & -02 21 39.6 &  12	$\pm$ 2 &  ... &  6 $\pm$ 1 &  0.97 $\pm$  0.42 &  -0.56	 &  -0.31\\
5 &  05 41  06.7 & -02 23 52.8 &  13	$\pm$ 2 &  5 $\pm$ 1 &  4 $\pm$ 1	 &  0.98 $\pm$  0.42 &  -0.71	 &  -1.00 \\
6 &  05 41 11.5 & -02 27 25.2 &  ... &  2 $\pm$ 1	 &  1	$\pm$ 1 &  0.22 $\pm$  0.44 &  ...	 &  ...\\
7 &  05 41 21.6 & -02 11  09.6 &  8 $\pm$ 1 &  ...	 &  ...	 &  0.93 $\pm$  0.41 &  -0.53	&  -0.15 & [YKK2000] A6\\
%%8 &  05 41 23.8 & -02 22 48.0 &  ...	 &  ...	 &  3 $\pm$ 1 &  0.34 $\pm$  0.44 &  -0.83 &  -1.00 & sp.  \\
%8 &  05 41 23.8 & -02 22 48.0 &  ... & ... & ... & ... & ... & ... & sp.  \\
%%9 &  05 41 23.8 & -02 03 28.8 &  13	$\pm$ 3 &  ...	 &  ...	 &  0.98 $\pm$  0.42 &  -0.62	 &  -0.05 & sp.  \\
%9 &  05 41 23.8 & -02 03 28.8 & ... & ... & ... & ... & ... & ... & sp.  \\
10 &  05 41 24.7 & -02 22 37.2 &  203 $\pm$ 6&  79 $\pm$ 4 &  77 $\pm$ 3 &  5.84 $\pm$ 1.28 &   0.24	 &  -0.49 & [YKK2000] A7\\
%%11 &  05 41 25.7 & -02 22  08.4 &  3	$\pm$ 1 &  ...	 &  ...	 &  0.87 $\pm$  0.41 &  -0.15 &  -1.00 & sp.  \\
%11 &  05 41 25.7 & -02 22  08.4 & ... & ... & ... & ... & ... & ... & sp.  \\
12 &  05 41 25.7 & -02 23  06.0 &  3	$\pm$ 1 &  ...	 &  ...	 &  0.87 $\pm$  0.41 &  -0.45 &  -0.34\\
13 &  05 41 28.3 & -02 13 58.8 &  4 $\pm$ 1&  ...	 &  ...	 &  0.88 $\pm$  0.41 &  -0.70 &  -1.00\\
14 &  05 41 28.3 & -02 19 44.4 &  3 $\pm$ 1&  ...	 &  ...	 &  0.87 $\pm$  0.41 &  -0.78 &  -1.00\\
15 &  05 41 31.9 & -02 10 19.2 &  5 $\pm$ 1&  ...	 &  ...	 &  0.89 $\pm$  0.41 &  -0.73 &  -1.00 &  \\
16 &  05 41 32.6 & -02 13 19.2 &  3 $\pm$ 1&  ...	 &  ...	 &  0.87 $\pm$  0.41 &  -1.00 &  -0.08\\
17 &  05 41 34.8 & -02 17 24.0 &  17	 $\pm$ 2 &  11 $\pm$ 1 &  9 $\pm$ 1 &  1.03 $\pm$  0.15 &  -0.20	 &  -0.89\\
18 &  05 41 36.5 & -02 16 48.0 &  ...	 &  1 $\pm$ 1 &  1 $\pm$ 1 &  0.22 $\pm$  0.44 &  -0.63	 &  -0.42\\
%%19 &  05 41 37.2 & -02 05 16.8 &  ...	 &  1 $\pm$ 1 &  ...	 &  0.55 $\pm$  0.39 &   ...	 &  -1.00 & sp.  \\
%19 &  05 41 37.2 & -02 05 16.8 & ... & ... & ... & ... & ... & ... & sp.  \\
20 &  05 41 43.0 & -02 10 26.4 &  100 $\pm$ 4&  26 $\pm$ 2&  27 $\pm$ 2 &  1.81 $\pm$ 0.22 &  -0.50	 &  -0.70 & [YKK2000] A9\\
21 &  05 41 43.2 & -02 17 38.4 &  2 $\pm$ 1&  ...	 &  ...	 &  0.86 $\pm$  0.41 &  -0.95 &  -1.00\\
22 &  05 41 43.2 & -02 03 54.0 &  161 $\pm$ 7&  51 $\pm$ 4&  48 $\pm$ 4 &  2.34 $\pm$ 2.30 &  -0.72	 &  -0.91 & [YKK2000] A10\\
23 &  05 41 43.7 & -02 10 22.8 &  ...	 &  4 $\pm$ 1&  ...	 &  0.72 $\pm$  0.39 &  -0.79 &  -1.00 & opt. source at $11\arcsec$\\
%23 &  05 41 43.7 & -02 10 22.8 & ... & ... & ... & ... & ... & ... & sp.  \\
24 &  05 41 44.2 & -02 16  08.4 &  3 $\pm$ 1&  ...	 &  ...	 &  0.87 $\pm$  0.41	 &  -0.95 &  -0.03\\
%%25 &  05 41 44.4 & -02 10 22.8 &  ...	 &  ...	 &  2 $\pm$ 1&  0.28 $\pm$  0.44	 &  -0.46 &  -1.00 & sp.  \\
%25 &  05 41 44.4 & -02 10 22.8 & ... & ... & ... & ... & ... & ... & sp.  \\
%%26 &  05 41 45.1 & -02 28 44.4 &  5 $\pm$ 1&  ...	 &  ...	 &  0.89 $\pm$  0.41	 &  -0.36	 &  -0.28 & sp.  \\
%26 &  05 41 45.1 & -02 28 44.4 & ... & ... & ... & ... & ... & ... & sp.  \\
27 &  05 41 45.6 & -02 24 18.0 &  14	 $\pm$ 2&  6 $\pm$ 1&  5 $\pm$ 1&  1.29 $\pm$  0.38 &   0.57	 &  -0.34\\
28 &  05 41 47.0 & -02 16 37.2 &  477 $\pm$ 7&  166 $\pm$ 4&  159 $\pm$ 4&  8.32 $\pm$ 0.87 &  -0.18	 &  -0.68 & [YKK2000] A11\\
%%29 &  05 41 47.8 & -02 16 40.8 &  ...	 &  ...	 &  10 $\pm$ 3 &  0.73 $\pm$  0.44	 &  -0.10	 &  -1.00 & sp.  \\
%29 &  05 41 47.8 & -02 16 40.8 & ... & ... & ... & ... & ... & ... & sp.  \\
%%30 &  05 41 47.8 & -02 17 31.2 &  3 $\pm$ 1 &  ...	 &  ...	 &  0.87 $\pm$  0.41	 &  -0.40	 &  -1.00 & sp.  \\
%30 &  05 41 47.8 & -02 17 31.2 & ... & ... & ... & ... & ... & ... & sp.  \\
31 &  05 41 48.2 & -02 10 40.8 &  7 $\pm$ 1 &  2 $\pm$ 1&  ...	 &  0.92 $\pm$  0.41	 &  -0.79	 &  -0.66\\
%%32 &  05 41 49.2 & -02 16 15.6 &  3 $\pm$ 1 &  ...	 &  ...	 &  1.02 $\pm$  0.47	 &   0.34	 &  -0.74 & sp.  \\
%32 &  05 41 49.2 & -02 16 15.6 & ... & ... & ... & ... & ... & ... & sp.  \\
33 &  05 41 51.1 & -02 29 52.8 &  ...	 &  15 $\pm$ 2 &  10 $\pm$ 2 &  0.73 $\pm$  0.44	 &  -0.01	 &  -0.70\\
34 &  05 41 52.6 & -02 03 50.4 &  131 $\pm$ 6 &  50 $\pm$ 4 &  35 $\pm$ 4 &  4.10 $\pm$  1.01	 &   0.30	 &  -0.61 & [YKK2000] A12\\
35 &  05 41 55.7 & -02 23 42.0 &  14	 $\pm$ 2 &  3 $\pm$ 1 &  ...	 &  1.26 $\pm$  0.47	 &   0.16	 &  -0.77\\
36 &  05 42  09.1 & -02 08 20.4 &  62 $\pm$ 6&  16 $\pm$ 2 &  20 $\pm$ 2 &  1.48 $\pm$  0.99 &  -0.05	 &  -0.55\\
%%37 &  05 42  09.4 & -02 23 16.8 &  ...	 &  1 $\pm$ 1 &  1 $\pm$ 1 &  0.22 $\pm$  0.44	 &  -0.05	 &  -0.18 & sp.  \\
%37 &  05 42  09.4 & -02 23 16.8 & ... & ... & ... & ... & ... & ... & sp.  \\
38 &  05 42 10.1 & -02 10 22.8 &  7 $\pm$ 1 &  ...	 &  ...	 &  0.92 $\pm$  0.41	 &  -0.91	 &  -1.00 &  \\
%%39 &  05 42 10.1 & -02 05 02.4 &  10	$\pm$ 3&  ...	 &  ...	 &  0.95 $\pm$  0.41	 &  -0.28	 &  -1.00 & sp.  \\
%39 &  05 42 10.1 & -02 05 02.4 & ... & ... & ... & ... & ... & ... & sp.  \\
%%40 &  05 42 10.3 & -02 04 44.4 &  ...	 &  5	$\pm$ 2 &  ...	 &  0.78 $\pm$  0.39	 &  -0.32	 &  -0.50 & sp.  \\
%40 &  05 42 10.3 & -02 04 44.4 & ... & ... & ... & ... & ... & ... & sp.  \\
41 &  05 42 12.5 & -02 05 09.6 &  517 $\pm$ 13 &  ...	 &  145 $\pm$ 7 &  9.90 $\pm$ 0.93 &  -0.36	 &  -0.53\\
42 &  05 42 16.1 & -02 06 46.8 &  72	 $\pm$ 5 &  27 $\pm$ 3 &  18 $\pm$ 2 &  1.57 $\pm$  0.74	 &  -0.71	 &  -0.61 & [YKK2000] A13\\
43 &  05 42 16.8 & -02 06 39.6 &  12	 $\pm$ 3&  6 $\pm$ 2 &  ...  	 &  0.97 $\pm$  0.42	 &  -0.89	 &  -0.43 & [YKK2000] A13\\
44 &  05 42 18.5 & -02 19 44.4 &  4 $\pm$ 1 &  ...	 &  ...	 &  0.88 $\pm$  0.41	 &  -0.95	 &  -0.58\\
45 &  05 42 19.7 & -02 15 32.4 &  ...	 &  1	$\pm$ 1&  3 $\pm$ 1 &  0.34 $\pm$  0.44	 &  ...	 &   ...\\
46 &  05 42 20.2 & -02 11 27.6 &  9 $\pm$ 2 &  3 $\pm$ 1 &  ...	 &  0.94 $\pm$  0.41	 &  -0.46	 &  -1.00\\
47 &  05 42 21.1 & -02 15 18.0 &  96 $\pm$ 4 &  34 $\pm$ 3 &  34 $\pm$ 2 &  1.77 $\pm$  0.91	 &  -0.06	 &  -0.68 & [YKK2000] A14?\\
48 &  05 42 21.4 & -02 07 44.4 &  50	 $\pm$ 4&  20 $\pm$ 3&  19 $\pm$ 2 &  1.36 $\pm$  0.59	 &  -0.38	 &  -0.92\\
49 &  05 42 33.8 & -02 09 46.8 &  9 $\pm$ 2 &  ...	 &  ...	 &  0.94 $\pm$  0.41	 &  ...	 &  ...\\
50 &  05 42 36.7 & -02 20 38.4 &  96	 $\pm$ 4&  25 $\pm$ 3&  35 $\pm$ 3 &  1.78 $\pm$ 1.58 &  -0.08	 &  -0.60 & \\
    \hline
  \end{tabular}
\flushleft{${(a)}$ Flux converted from observed count-rates as explained in Section~\ref{discussion};
${(b)}$ As in Tables~\ref{tabNIR} and \ref{tabIR}, we have removed sources considered as spurious. 
[YKK2000] is for \citet{yam00}.}

\end{table*}

\begin{onecolumn}
%\begin{landscape}
\begin{table*}
\begin{minipage}{\linewidth}
%\tiny
\caption{Optical and near-infrared magnitudes for the sources detected in the \textit{XMM-Newton} observation.}
\label{tabNIR}
\small
\centering
\begin{tabular}{ccrrrrrrl}
\hline
 Src. ID & Name & \multicolumn{1}{c} {V} & \multicolumn{1}{c} {I} & \multicolumn{1}{c} {J} & \multicolumn{1}{c} {H} & \multicolumn{1}{c} {K$_\mathrm{s}$} & \multicolumn{1}{c} {A$_\mathrm{V}$} & \multicolumn{1}{l} {IR-Class$^\dag$ and}\\
       & & \multicolumn{1}{c} {(mag)} & \multicolumn{1}{c} {(mag)} & \multicolumn{1}{c} {(mag)} & \multicolumn{1}{c} {(mag)} & \multicolumn{1}{c} {(mag)} & \multicolumn{1}{c} {(mag)} & \multicolumn{1}{l} {other remarks} \\
\hline
1  &  &  ...  &  14.06 $\pm$ 0.03  & 12.81 $\pm$ 0.03  & 12.11 $\pm$ 0.02  &  11.93 $\pm$ 0.02  & 1.0 $\pm$ 0.5 &  III \\ 
%2  &  & ...  &  ... & ...  & ...  &  ...  &  sp. \\
3  & HD 37805/MIR 31  & 7.53 $\pm$ 0.05  &  7.23 $\pm$ 0.05  & 6.95 $\pm$ 0.03  & 6.87 $\pm$ 0.05  &  6.76 $\pm$ 0.02  & 0.0 $\pm$ 0.5 &  III \\ 
4  &  & ...  & 14.48 $\pm$ 0.03  & 12.86 $\pm$ 0.03  & 12.24 $\pm$ 0.03  &  11.95 $\pm$ 0.02  & 0.7 $\pm$ 0.7 &  III \\ 
5  &  & ...  & 14.56 $\pm$ 0.03  & 13.03 $\pm$ 0.03  & 12.45 $\pm$ 0.03  &  12.21 $\pm$ 0.02  & 0.0 $\pm$ 0.5 & III \\
6  &  & ...  & 17.31 $\pm$ 0.11  & 13.70 $\pm$ 0.08  & ...  &  11.34 $\pm$ 0.08 &  ... & III \\ 
7  & VLA 3/MIR 46 & ... &  ...  & ...  & ...  &  ...  &  ... & I \\ 
%8  &  & ...  &  ...  & ...  & ...  &  ...  &  sp. \\
%9  &  & ...  &  ...  & ...  & ...  &  ...  &  sp. \\
10  & V615 Ori/MIR 52  & 13.11 $\pm$ 0.24  &  10.99 $\pm$ 0.17  & 9.10 $\pm$ 0.02 & 8.07 $\pm$ 0.04  &  7.39 $\pm$ 0.02  & 12.2 $\pm$ 3.5 &  II \\ 
%11  &  & ...  &  ...  & ...  & ...  &  ...  &  sp. \\
12  &  & ...  &  ...  & ...  & ...  &  ...  &  ... & - \\ 
13  &  & ...  &  14.21 $\pm$ 0.03  & 12.09 $\pm$ 0.02  & 11.12 $\pm$ 0.02  &  10.73 $\pm$ 0.02  & 6.6 $\pm$ 1.5 &  I/II \\%Trans. Disk? \\
14  & MIR 55  & 12.83 $\pm$ 0.25  &  12.03 $\pm$ 0.09  & 11.34 $\pm$ 0.02  & 10.94 $\pm$ 0.02  &  10.89 $\pm$ 0.02  & 0.0 $\pm$ 0.6 &   III \\ 
15  &  & ...  &  15.74 $\pm$ 0.05  & 14.23 $\pm$ 0.03  & 13.60 $\pm$ 0.03  & 13.33 $\pm$ 0.04  & 0.7 $\pm$ 0.5 &  III \\
16  & NIR 17  & ...  &  ...  & 17.29 $\pm$ 0.23  & 15.55 $\pm$ 0.10  &  14.47 $\pm$ 0.08  & 23.8 $\pm$ 6.0 &  III? \\
17  & MIR 60  & ...  &  14.27 $\pm$ 0.03  & 12.28 $\pm$ 0.03  & 11.42 $\pm$ 0.02 &  10.97 $\pm$ 0.02  & 1.0 $\pm$ 0.5 &  I/II \\ 
18  & MIR 62  & ...  &  14.75 $\pm$ 0.03  & 11.73 $\pm$ 0.03  & 10.05 $\pm$ 0.02 &  8.83 $\pm$ 0.02  & 8.5 $\pm$ 1.5 &  II (AGN?) \\ 
%19  &  & ...  &  ...  & ...  & ...  &  ...  &  sp. \\
20  & MIR 71  & 13.58 $\pm$ 0.39  &  12.28 $\pm$ 0.10  & 11.25 $\pm$ 0.02  & 10.63 $\pm$ 0.02  &  10.49 $\pm$ 0.02  & 0.0 $\pm$ 0.5 &  III \\ 
21  & WB 57/NIR 24  & ...  &  14.57 $\pm$ 0.03  & 13.03 $\pm$ 0.03  & 12.35 $\pm$ 0.02 &  12.11 $\pm$ 0.03  & 1.0 $\pm$ 0.3 &  III \\%Trans. Disk? (bad ch 3 and 4 photometry) \\ 
22  &  & 13.13 $\pm$ 0.24  &  11.83 $\pm$ 0.10  & 10.93 $\pm$ 0.03 & 10.32 $\pm$ 0.02  &  10.16 $\pm$ 0.02  & 0.0 $\pm$ 0.8 &  III \\ 
23  &  & ...  &  ...  & ...  & ...  &  ...  &  ... & opt. source at $11\arcsec$ \\
24  & W 218/NIR 25  & ...  &  ...  & 13.38 $\pm$ 0.04  & 12.51 $\pm$ 0.05  & 11.93 $\pm$ 0.04  & ... & not reliable$^\ddag$ \\ 
%24  & W 218/NIR 25  & ...  &  ...  & ...  & ...  & ...  & - \\
%25  &  & ...  &  ...  & ...  & ...  &  ...  &  sp. \\
%26  &  & ...  &  ...  & ...  & ...  &  ...  &  sp. \\
27  & MIR 76  & ...  &  16.90 $\pm$ 0.11  & 13.19 $\pm$ 0.03  & 11.52 $\pm$ 0.02 &  10.80 $\pm$ 0.02  & 14.8 $\pm$ 1.5 &  III \\ 
28  & MIR 80  & 12.39 $\pm$ 0.20  &  10.64 $\pm$ 0.13  & 9.04 $\pm$ 0.02  & 8.14 $\pm$ 0.04  &  7.55 $\pm$ 0.03  & 1.0 $\pm$ 1.0 &  II \\ 
%29  &  & ...  &  ...  & ...  & ...  &  ...  &  sp. \\
%30  &  & ...  &  ...  & ...  & ...  &  ...  &  sp. \\
31  & MIR 81  & 11.93 $\pm$ 0.14  &  11.30 $\pm$ 0.07  & 10.89 $\pm$ 0.03  & 10.62 $\pm$ 0.02  &  10.58 $\pm$ 0.02  & 0.5 $\pm$ 0.5 &  II/III \\ %(bad 8 microns) \\ 
%32  &  & ...  &  ...  & ...  & ...  &  ...  &  sp. \\
33  & MIR 83  & ...  &  10.00 $\pm$ 0.22  & 8.90 $\pm$ 0.02  & 8.60 $\pm$ 0.02 &  8.45 $\pm$ 0.02  & 2.9 $\pm$ 0.5 &  III \\ 
34  & V621 Ori  & ...  &  12.63 $\pm$ 0.02  & 10.57 $\pm$ 0.03  & 9.51 $\pm$ 0.02  &  9.11 $\pm$ 0.02  & 7.5 $\pm$ 0.7 &  III \\ 
35  & MIR 86  & ...  &  15.76 $\pm$ 0.07  & 12.85 $\pm$ 0.02  & 11.66 $\pm$ 0.02 &  11.05 $\pm$ 0.02  & 5.8 $\pm$ 0.5 &  II (AGN?) \\ 
36  &  & ...  &  12.74 $\pm$ 0.03  & 11.07 $\pm$ 0.02  & 10.15 $\pm$ 0.02  &  9.87 $\pm$ 0.02  & 1.7 $\pm$ 0.5 &  II/III \\ 
%37  &  & ...  &  ...  & ...  & ...  &  ...  &  sp. \\
38  &  & 12.60 $\pm$ 0.20  &  11.88 $\pm$ 0.08  & 11.31 $\pm$ 0.02  & 10.97 $\pm$ 0.02  &  10.93 $\pm$ 0.02  & 0.0 $\pm$ 1.5 &  III \\ 
%39  &  & ...  &  ...  & ...  & ...  &  ...  &  sp. \\
%40  &  & ...  &  ...  & ...  & ...  &  ...  &  sp. \\
41  &  & ...  &  13.36 $\pm$ 0.03  & 11.84 $\pm$ 0.02  & 11.15 $\pm$ 0.02  & 10.92 $\pm$ 0.02  & 1.0 $\pm$ 0.5 &  II/III \\ 
42  & V622 Ori  & 12.96 $\pm$ 0.23  &  11.75 $\pm$ 0.12  & 10.95 $\pm$ 0.02  & 10.43 $\pm$ 0.02  &  10.31 $\pm$ 0.02  & 1.5 $\pm$ 2.0 &  III \\ 
43  &  & ...  &  13.76 $\pm$ 0.03  & 12.21 $\pm$ 0.02  & 11.51 $\pm$ 0.03  & 11.21 $\pm$ 0.02  & 0.0 $\pm$ 1.5 &  II/III \\ 
44  &  & ...  &  14.29 $\pm$ 0.04  & 12.79 $\pm$ 0.03  & 12.06 $\pm$ 0.02  & 11.85 $\pm$ 0.03  & 1.5 $\pm$ 0.7 &  II/III \\ 
45  & V623 Ori  & ...  &  14.10 $\pm$ 0.04  & 12.80 $\pm$ 0.08  & ...  & 11.53 $\pm$ 0.10  & ... &  II/III \\
46  &  & ...  &  15.54 $\pm$ 0.06  & 13.73 $\pm$ 0.02  & 13.17 $\pm$ 0.03  & 12.92 $\pm$ 0.03  & 0.5 $\pm$ 0.6 &  II \\
47  &  & 13.03 $\pm$ 0.18  &  11.28 $\pm$ 0.14  & 9.40 $\pm$ 0.02  & 8.68 $\pm$ 0.04  &  8.39 $\pm$ 0.02  & 0.5 $\pm$ 0.5 &  III \\ 
48  &  & ...  &  13.67 $\pm$ 0.03  & 12.28 $\pm$ 0.02  & 11.62 $\pm$ 0.02  & 11.37 $\pm$ 0.02  & 1.0 $\pm$ 0.6 &  III \\
49  &  & ...  &  14.48 $\pm$ 0.04  & 12.57 $\pm$ 0.02  & 12.00 $\pm$ 0.03  & 11.67 $\pm$ 0.02  & 0.0 $\pm$ 0.7 &  III \\ 
50  &  & ...  &  13.32 $\pm$ 0.03  & 11.56 $\pm$ 0.03  & 10.68 $\pm$ 0.02  & 10.39 $\pm$ 0.02  & 4.5 $\pm$ 1.0 &  III \\ 
\hline
\end{tabular}
\\
%\flushleft{* IRAC flux upper limits for X-ray sources with signal above the detection limit in the \textit{XMM-Newton} observation.}
\flushleft{$^\dag$ Infrared object class from the SED and Spitzer color-color diagrams and remarks for particular sources. We have removed spurious sources (2, 8, 9, 11, 19, 25, 26, 29, 30, 32, 37, 39, 40). $^\ddag$ The photometry of Src. 24 is not reliable due to confusion of sources in the region.}
\end{minipage}
\end{table*}
%\end{landscape}
\end{onecolumn}

\begin{onecolumn}
\begin{landscape}
\begin{table*}
\begin{minipage}{\linewidth}
%\tiny
\caption{Infrared photometric data for the sources detected in the \textit{XMM-Newton} observation. $W$ is for \textit{WISE}, $S$ is for \textit{Spitzer}. }
\label{tabIR}
\scriptsize
\centering
\begin{tabular}{ccrrrrrrrrrl}
\hline
 Src. ID & Name & \multicolumn{1}{c}{F$^W_{3.4}$ (mJy)} & \multicolumn{1}{c}{F$^S_{3.6}$ (mJy)} & \multicolumn{1}{c}{F$^S_{4.5}$ (mJy)} & \multicolumn{1}{c}{F$^W_{4.6}$ (mJy)} & \multicolumn{1}{c}{F$^S_{5.4}$ (mJy)} & \multicolumn{1}{c}{F$^S_{8.0}$(mJy)} &  \multicolumn{1}{c}{F$^W_{12}$ (mJy)} & \multicolumn{1}{c}{F$^W_{22}$ (mJy)} &  
\multicolumn{1}{c}{F$^S_{24}$ (mJy)} &  IR class$^\dag$/Remarks\\
%
%       & &  \multicolumn{1}{c}{(mJy)} & \multicolumn{1}{c}{(mJy)} & \multicolumn{1}{c}{(mJy)} & 
    %   \multicolumn{1}{c}{(mJy)} &  \multicolumn{1}{c}{(mJy)} & \multicolumn{1}{c}{(mJy)} & \multicolumn{1}{c}{(mJy)} & 
       %\multicolumn{1}{c}{(mJy)} & \multicolumn{1}{c}{(mJy)} & other remarks\\
\hline
1  &  & 5.9 $\pm$ 0.4  &  5.3 $\pm$ 0.3  & 3.5 $\pm$ 0.3  & 3.5 $\pm$ 0.3  &  1.6 $\pm$ 0.6  &  1.0 $\pm$ 0.1 &  ...  &  ...  &  ...  &  III \\ 
%2  &  & ...  &  ...  & ...  & ...  &  ...  &  ...  &  ...  &  ...  &  ...  &  sp. \\
3  & HD 37805/MIR 31  & ...  &  518.2 $\pm$ 5.2  & 336.5 $\pm$ 3.5  & ...  &  219.7 $\pm$ 2.5  &  149.8 $\pm$ 2.0  &  ...  &  ...  &  55.4 $\pm$ 0.2  & III \\ 
4  &  & 5.5 $\pm$ 0.4  &  5.1 $\pm$ 0.3  & 4.6 $\pm$ 0.3  & 3.8 $\pm$ 0.3  & ...  &  ...  &  ...  &  ...  &  ...  &  III \\ 
5  &  & ...  &  3.6 $\pm$ 0.3  & 3.1 $\pm$ 0.3  & ...  &  ...  &  ...  &  ... &  ...  & ...  &  III \\
6  &  & 13.1 $\pm$ 0.5  &  ...  & 8.2 $\pm$ 0.3  & ...  &  ...  &  ...  &  ... &  ...  & ...  &  III \\ 
7  & VLA 3/MIR 46 & ...  &  3.0 $\pm$ 0.2  & 4.6 $\pm$ 0.3  & ...  & 6.5 $\pm$ 0.5  &  13.6 $\pm$ 0.6  &  ...  &  ...  &  23.4 $\pm$ 0.2  &  I \\ 
%8  &  & ...  &  ...  & ...  & ...  &  ...  &  ...  &  ...  &  ...  &  ...  & sp. \\
%9  &  & ...  &  ...  & ...  & ...  &  ...  &  ...  &  ...  &  ...  &  ...  & sp. \\
10  & V615 Ori/MIR 52  & 998.9 $\pm$ 36.3  &  931.9 $\pm$ 9.0  & 995.2 $\pm$ 9.6 & 1468.9 $\pm$ 48.3  &  1020.1 $\pm$ 9.9  &  1207.5 $\pm$ 11.7  & 1215.3 $\pm$ 32.9  &  1396.89 $\pm$ 55.5  &  1067.0 $\pm$ 0.5  &  II \\ 
%11  &  & ...  &  ...  & ...  & ...  &  ...  &  ...  &  ...  &  ...  &  ...  & sp. \\
12  &  & ...  &  0.1 $\pm$ 0.2  & 0.7 $\pm$ 0.3  & ...  &  2.2 $\pm$ 0.7  & 6.1 $\pm$ 1.0  &  ...  &  ...  &  ...  &  - \\ 
13  &  & 23.6 $\pm$ 1.3  &  20.9 $\pm$ 0.5  & 11.8 $\pm$ 0.4  & 16.1 $\pm$ 1.2 &  27.3 $\pm$ 1.0  &  ...  &  ...  &  ...  &  ..  &  I/II \\%Trans. Disk? \\
14  & MIR 55  & 14.0 $\pm$ 0.4  &  12.4 $\pm$ 0.4  & 7.4 $\pm$ 0.3  & 7.5 $\pm$ 0.2  &  9.3 $\pm$ 0.6  &  4.5 $\pm$ 0.4  &  ...  &  ...  &  ...  &  III \\ 
15  &  & 1.5 $\pm$ 0.1  &  1.1 $\pm$ 0.2  & 1.3 $\pm$ 0.2  & 1.0 $\pm$ 0.1  & ...  &  ...  &  ...  &  ...  &  ...  &  III \\
16  & NIR 17  & ...  &  ...  & ...  & ...  &  ...  &  ...  &  ...  &  ...  & ...  &  III? \\
17  & MIR 60  & 21.1 $\pm$ 0.8  &  20.7 $\pm$ 0.5  & 15.8 $\pm$ 0.5  & 19.59 $\pm$ 0.7  &  40.4 $\pm$ 2.1  &  62.5 $\pm$ 2.4  &  156.1 $\pm$ 18.4  & 559.7 $\pm$ 79.3  &  ...  &  I/II \\ 
18 & MIR 62 & 508.1 $\pm$ 21.9 & 595.6 $\pm$ 6.0 & 686.7 $\pm$ 6.8 & 700.5 $\pm$ 20.8 & 764.1 $\pm$ 10.0 & 957.6 $\pm$ 21.8 & ... & 2560.7 $\pm$ 162.5 & 1063.0 $\pm$ 1.0 & II  (AGN?) \\ 
%19  &  & ...  &  ...  & ...   & ...  &  ... & ...  &  ...  &  ...  &  ...  & sp. \\
20  & MIR 71  & ...  &  19.0 $\pm$ 0.5  & 12.4 $\pm$ 0.4  & ...  & 7.4 $\pm$ 0.6  &  4.9 $\pm$ 0.6  &  ...  &  ...  &  ...  &  III \\ 
21  & WB 57/NIR 24  & ...  &  5.6 $\pm$ 0.3  & 3.0 $\pm$ 0.3  & ...  &  ...  &  ...  & ...  &  ...  &  ...  &  III \\%Trans. Disk? (bad ch 3 and 4 photometry) \\ 
22  &  & 28.7 $\pm$ 0.7  &  25.4 $\pm$ 0.6  & ...  & 16.24 $\pm$ 0.4  & 11.6 $\pm$ 0.6  &  ...  &  ...  &  ...  &  ...  &  III \\ 
23  &  & ...  &  ...  & ...  & ...  &  ...  &  ...  &  ...  &  ...  &  ...  & opt. source at $11\arcsec$ \\
24  & W 218/NIR 25  & ...  &  295.4 $\pm$ 3.2  & 501.3 $\pm$ 5.1  & ...  & 706.4 $\pm$ 7.3  &  1161.2 $\pm$ 12.6  &  ...  &  ...  &  ...  &  not reliable$^\ddag$ \\ 
%24 & W 218/NIR 25  & ...  & ...  & ...  & ...  & ...  & ...  & ...  & ...  & ...  & - \\
%25  &  & ...  &  ...  & ...  & ...  &  ...  &  ...  &  ...  &  ...  &  ...  & sp. \\
%26  &  & ...  &  ...  & ...  & ...  &  ...  &  ...  &  ...  &  ...  &  ...  & sp. \\
27  & MIR 76  & 19.7 $\pm$ 0.5  &  20.3 $\pm$ 0.5  & 13.9 $\pm$ 0.4  &14.7 $\pm$ 0.4  &  13.2 $\pm$ 0.7  &  6.0 $\pm$ 0.3  &  ...  &  ...  &  ...  & III \\ 
28  & MIR 80  & 811.2 $\pm$ 30.8  &  785.9 $\pm$ 7.7  & 755.0 $\pm$ 7.4  & 944.0 $\pm$ 27.2  &  719.1 $\pm$ 7.2  &  789.3 $\pm$ 8.7  &  827.8 $\pm$ 21.1 &  785.5 $\pm$ 33.2  &  458.7 $\pm$ 0.3  &  II \\ 
%29  &  & ...  &  ...  & ...  & ...  &  ...  &  ...  &  ...  &  ...  &  ...  & sp. \\
%30  &  & ...  &  ...  & ...  & ...  &  ...  &  ...  &  ...  &  ...  &  ...  & sp. \\
31  & MIR 81  & 19.1 $\pm$ 0.5  &  14.9 $\pm$ 0.4  & 10.3 $\pm$ 0.4  & 10.37 $\pm$ 0.3  &  5.3 $\pm$ 0.6  &  ...  &  4.3 $\pm$ 1.6  &  ...  &  ...  & II/III \\ %(bad 8 microns) \\ 
%32  &  & ...  &  ...  & ...  & ...  &  ...  &  ...  &  ...  &  ...  &  ...  & sp. \\
33  & MIR 83  & 160.5 $\pm$ 4.3  &  134.1 $\pm$ 1.7  & 89.7 $\pm$ 1.2  & 91.1 $\pm$ 2.3  &  65.9 $\pm$ 1.2  &  36.0 $\pm$ 0.7  &  11.6 $\pm$ 1.5  & ...  &  24.4 $\pm$ 0.2  &  III \\ 
34  & V621 Ori  & 83.8 $\pm$ 2.2  &  74.7 $\pm$ 1.1  & ...  & 50.8 $\pm$ 1.3  & 38.2 $\pm$ 0.8  &  ...  &  12.4 $\pm$ 0.5  &  ...  &  ...  &  III \\ 
35  & MIR 86  & 22.6 $\pm$ 0.5  &  27.2 $\pm$ 0.6  & 27.1 $\pm$ 0.6  & 24.1 $\pm$ 0.6  &  30.5 $\pm$ 0.8  &  38.9 $\pm$ 0.8  &  58.3 $\pm$ 2.2  & 154.9 $\pm$ 9.6  &  ...  &  II (AGN?) \\ 
36  &  & 39.8 $\pm$ 1.1  &  33.2 $\pm$ 0.6  & 21.0 $\pm$ 0.5  & 22.1 $\pm$ 0.5 &  16.1 $\pm$ 0.6  &  11.6 $\pm$ 0.5  &  12.8 $\pm$ 1.0  &  17.9 $\pm$ 4.5  & ...  &  II/III \\ 
%37  &  & ...  &  ...  & ...  & ...  &  ...  &  ...  &  ...  &  ...  &  ...  & sp. \\
38  &  & 13.9 $\pm$ 0.4  &  11.0 $\pm$ 0.4  & 7.8 $\pm$ 0.3  & 7.6 $\pm$ 0.2  & 8.1 $\pm$ 0.6  &  3.5 $\pm$ 0.3  &  ...  &  ...  &  ...  &  III \\ 
%39  &  & ...  &  ...  & ...  & ...  &  ...  &  ...  &  ...  &  ...  &  ...  & sp. \\
%40  &  & ...  &  ...  & ...  & ...  &  ...  &  ...  &  ...  &  ...  &  ...  & sp. \\
41  &  & 15.4 $\pm$ 0.4  &  13.3 $\pm$ 0.4  & ...  & 9.4 $\pm$ 0.2  & 6.9 $\pm$ 0.7  &  ...  &  3.4 $\pm$ 0.6  &  ...  &  ...  &  II/III \\ 
42  & V622 Ori  & 24.5 $\pm$ 0.7  &  20.4 $\pm$ 0.5  & 13.0 $\pm$ 0.4  & 13.4 $\pm$ 0.3  &  6.3 $\pm$ 0.6  &  6.8 $\pm$ 0.3  &  4.8 $\pm$ 0.9  &  ... &  ..  &  III \\ 
43  &  & 13.2 $\pm$ 0.4  &  11.1 $\pm$ 0.4  & 9.2 $\pm$ 0.4  & 9.6 $\pm$ 0.2  & 8.7 $\pm$ 0.6  &  7.5 $\pm$ 0.4  &  4.6 $\pm$ 0.1  &  ...  &  ...  & II/III \\ 
44  &  & 6.3 $\pm$ 0.2  &  5.1 $\pm$ 0.3  & 3.7 $\pm$ 0.3  & 3.9 $\pm$ 0.1  & 3.4 $\pm$ 0.5  &  2.1 $\pm$ 0.3  &  2.6 $\pm$ 0.8  &  5.7 $\pm$ 1.6 $\pm$ & ...  &  II/III \\ 
45  & V623 Ori  & 9.1 $\pm$ 0.2  &  ...  & ...  & 8.8 $\pm$ 0.2  &  ...  &  ... &  7.6 $\pm$ 0.9  &  13.7 $\pm$ 4.2  &  ...  &  II/III \\
46  &  & 2.5 $\pm$ 0.1  &  2.3 $\pm$ 0.2  & 2.0 $\pm$ 0.2  & 1.8 $\pm$ 0.1  & 0.9 $\pm$ 0.5  &  ...  &  2.8 $\pm$ 0.7  &  9.9 $\pm$ 2.7 $\pm$ &  ...  & II \\
47  &  & 149.9 $\pm$ 4.0  &  131.1 $\pm$ 1.6  & 79.7 $\pm$ 1.1  & 85.4 $\pm$ 2.1  &  61.0 $\pm$ 1.1  &  35.6 $\pm$ 0.7  &  19.8 $\pm$ 1.1  & 9.6 $\pm$ 3.2  &  ...  &  III \\ 
48  &  & 8.8 $\pm$ 0.2  &  9.5 $\pm$ 0.3  & 6.0 $\pm$ 0.3  & 5.6 $\pm$ 0.2  & 3.1 $\pm$ 0.6  &  ...  &  ...  &  ...  &  ...  &  III \\
49  &  & 7.4 $\pm$ 0.2  &  7.9 $\pm$ 0.3  & 5.3 $\pm$ 0.3  & 5.1 $\pm$ 0.1  & 3.0 $\pm$ 0.5  &  1.8 $\pm$ 0.3  &  ...  &  ...  &  ...  &  III \\ 
50  &  & 24.8 $\pm$ 0.7  &  22.1 $\pm$ 0.5  & 15.3 $\pm$ 0.5  & 14.6 $\pm$ 0.4 &  11.9 $\pm$ 0.6  &  5.3 $\pm$ 0.4  &  ...  &  ...  &  ...  &  III \\
\hline
\end{tabular}
\\
%\flushleft{* IRAC flux upper limits for X-ray sources with signal above the detection limit in the \textit{XMM-Newton} observation.}
\flushleft{$^\dag$ Infrared object class from the SED and Spitzer color-color diagrams and remarks for particular sources. We have removed spurious sources (2, 8, 9, 11, 19, 25, 26, 29, 30, 32, 37, 39, 40). $^\ddag$ The photometry of Src. 24 is not reliable due to confusion of sources in the region.}
\end{minipage}
\end{table*}
\end{landscape}
\end{onecolumn}

\begin{onecolumn}
%\begin{landscape}
\begin{table*}
\begin{minipage}{\linewidth}
%\tiny
\caption{Main results from the use of the fitting tool of \citet{rob07} with the NGC 2023 X-ray detected sources.}
\label{tabSEDs}
\small
\centering
\begin{tabular}{cccccccccccc}
\hline
 Src. ID & Name & \multicolumn{1}{c} {A$_\mathrm{V}$} & \multicolumn{1}{c}{Distance} & \multicolumn{1}{c} {$M_\star$} & \multicolumn{1}{c} {$R_\star$} & \multicolumn{1}{c} {$T_\star$} & \multicolumn{1}{c} {$L_\mathrm{bol}$} & \multicolumn{1}{c} {$\log L_\mathrm{X}/L_\mathrm{bol}$} & \multicolumn{1}{c}{$\dot{M}$} &  Stage$^{a}$ & IR Class$^{b}$\\
       & & \multicolumn{1}{c} {(mag)} & \multicolumn{1}{c} {(pc)} & \multicolumn{1}{c}{(M$_\odot$)} & \multicolumn{1}{c}{(R$_\odot$)} & \multicolumn{1}{c}{(K)} & \multicolumn{1}{c} {(L$_\odot$)} & & 
       \multicolumn{1}{c} {(M$_\odot/yr$)} \\
\hline
%1   & & 1.04 & 355 & 0.33 & 0.47 & 3500 & 0.01 & ~-3.27 & ... & ... & III \\ 
1   & & 0.05 & 398 & 0.48 & 1.02 & 3735 & 0.19 & ~-2.68 & ... & ... & III \\ 
3   & HD 37805/MIR 31 &  0.10 & 56 & 1.45 & 1.77 & 7500 & 8.53 & -13.02 & ... & ... & III \\
4   & & 1.11 & 299 & 0.35 & 0.77 & 3550 & 0.09 & ~-3.41 & ... & ... & III \\
5   & & 0.25 & 398 & 0.38 & 0.52 & 3650 & 0.01 & ~-3.08 & ... & ... & III \\
6   & & 8.81 & 630 & 0.66 & 0.76 & 4672 & 0.27 & ~-3.59 & ... & ...  & III \\
10 & V615 Ori/ MIR 52 & 5.19 & 371 & 2.96 & 4.39 & 7902 & 67.4 & -69.89 & $1.82 \times 10^{-9}$ & III & II \\
13 & & 1.90 & 500 & 1.09 & 10.14 & 4037 & 24.6 & -27.52 & $1.03 \times 10^{-8}$ & I & I/II \\
%14 & MIR 55 & 0.00 & 575 & 0.76 & 0.84 & 5083 & 0.38 & ~-3.18 & ... & ... & III \\
14 & MIR 55 & 0.00 & 575 & 1.60 & 1.88 & 5084 & 2.11 & ~-3.37 & ... & ... & III \\
%15 & & 0.70 & 195 & 0.29 & 0.42 & 3368 & 0.05 & ~-3.78 & ... & ...  & III \\
15 & & 0.24 & 479 & 0.29 & 0.76 & 3399 & 0.07 & ~-3.78 & ... & ...  & III \\
17 & MIR 60 & 1.19 & 575 & 0.48 & 5.40 & 3692 & 4.87 & ~-8.36 & $2.15 \times 10^{-9}$ & I & I/II \\
18 & MIR 62 & 5.33 & 457 & 3.42 & 2.63 & 11725 & 118 & -121.6 & $4.51 \times 10^{-9}$ & III & II \\
%20 & MIR 71 & 0.76 & 347 & 0.69 & 0.79 & 4809 & 0.31 & ~-3.40 & ... & ... & III \\
20 & MIR 71 & 0.76 & 347 & 1.31 & 1.46 & 4809 & 1.02 & ~-3.18 & ... & ... & III \\
%21 & NIR 24 & 0.42 & 245 & 0.29 & 0.41 & 3365 & 0.02 & ~-3.57 & ... & ... & III \\
21 & NIR 24 & 0.42 & 245 & 0.24 & 0.72 & 3366 & 0.06 & ~-2.57 & ... & ... & III \\
%22 & & 0.30 & 280 & 0.47 & 0.63 & 3900 & 0.06 & -3.15 & ... & ... & III \\
22 & & 0.34 & 288 & 1.35 & 1.52 & 4886 & 1.18 & -3.29 & ... & ... & III \\
%24 & NIR 25 & 5.30 & 210 & 5.44 & 12.0 & 6653 & 255 & $6.44 \times 10^{-10}$ & II \\
%27 & MIR 76 & 7.95 & 250 & 0.31 & 0.30 & 3555 & 0.02 & ~-3.61 & ... & ... & III \\
27 & MIR 76 & 7.65 & 275 & 0.35 & 1.92 & 3500 & 0.63 & ~-3.30 & ... & ... & III \\
28 & MIR 80 & 3.11 & 436 & 2.96 & 4.39 & 7902 & 67.4 & -69.09 & $1.82 \times 10^{-9}$ & III & II \\
31 & MIR 81 & 0.88 & 616 & 1.85 & 1.53 & 7978 & 9.32 & -12.03 & $1.70 \times 10^{-14}$ & III & II/III \\
%33 & MIR 83 & 2.15 & 280 & 1.46 & 1.28 & 6835 & 3.97 & ~-7.47 & ... & ... & III \\
33 & MIR 83 & 2.15 & 280 & 1.93 & 2.78 & 6835 & 15.1 & ~-4.92 & ... & ... & III \\
%34 & V621 Ori & 4.30 & 310 & 0.70 & 0.79 & 4828 & 0.31 & ~-3.00 & ... & ... & III \\
34 & V621 Ori & 4.30 & 309 & 2.05 & 2.96 & 4828 & 4.27 & ~-3.54 & ... & ... & III \\
35 & MIR 86 & 3.87 & 550 & 0.22 & 2.86 & 3119 & 0.93 & ~-3.61 & $1.07 \times 10^{-7}$ & I & II \\
%36 & & 2.88 & 309 & 0.85 & 0.76 & 4847 & 0.32 & ~-3.51 & $2.70 \times 10^{-13}$ & III & II/III \\
36 & & 2.88 & 309 & 1.60 & 1.91 & 4847 & 1.82 & ~-3.61 & $2.70 \times 10^{-13}$ & III & II/III \\
38 & & 0.07 & 199 & 0.33 & 0.47 & 3500 & 0.01 & ~-3.71 & ... & ... & III \\  % No tocada en la segunda revision con los resultados del modelo de Robitaille. 
41 & & 0.55 & 166 & 0.35 & 0.77 & 3567 & 0.09 & ~-3.08 & $2.54 \times 10^{-10}$ & III & II/III \\
%42 & V622 Ori & 0.11 & 407 & 0.82 & 0.71 & 4759 & 0.23 & ~-3.08 & $1.13 \times 10^{-13}$ & III & II/III \\
42 & V622 Ori & 0.11 & 407 & 1.52 & 1.83 & 4759 & 1.54 & ~-3.27 & $1.13 \times 10^{-13}$ & III & III \\
43 & & 0.37 & 182 & 0.24 & 0.82 & 3343 & 0.08 & ~-3.84 & $1.53 \times 10^{-11}$ & III & II/III \\
44 & & 0.51 & 257 & 0.31 & 0.29 & 3556 & 0.02 & ~-3.52 & $7.88 \times 10^{-12}$ & III & II/III \\
45 & V623 Ori& 0.13 & 309 & 0.26 & 0.27 & 3401 & 0.01 & ~-3.76 & $1.09 \times 10^{-8}$ & III & II/III \\
46 & & 0.07 & 500 & 0.13 & 0.15 & 3052 & 0.002 & ~-2.89 & $3.07 \times 10^{-11}$ & III & II \\
47 & & 1.49 & 148 & 0.73 & 0.61 & 4331 & 0.13 & ~-3.80 & $3.00 \times 10^{-14}$ & III & III \\  % No tocada en la segunda revision con los resultados del modelo de Robitaille. 
%48 & & 0.37 & 280 & 0.45 & 0.60 & 3850 & 0.06 & ~-3.29 & ... & ... & III \\
48 & & 0.09 & 275 & 0.36 & 1.00 & 3548 & 1.43 & ~-3.64 & ... & ... & III \\
%49 & & 0.49 & 263 & 0.19 & 0.29 & 3047 & 0.002 & ~-3.45 & ... & ... & III \\
49 & & 0.29 & 398 & 0.17 & 0.16 & 3042 & 0.20 & ~-2.63 & ... & ... & III \\
%50 & & 2.92 & 199 & 0.68 & 0.78 & 4750 & 0.29 & ~-3.68 & ... & ... & III \\
50 & & 1.37 & 190 & 0.46 & 1.11 & 3706 & 0.21 & ~-3.02 & ... & ... & III \\
\hline
\end{tabular}
\\
\flushleft{$(a)$ Evolutionary stage parameter from \citet{rob07}. $(b)$ Infrared class based on the SEDs 
(Tables~\ref{tabNIR} and \ref{tabIR}).}
\end{minipage}
\end{table*}
%\end{landscape}
\end{onecolumn}

%_____________________________________________________________
 \begin{figure*}
 \centering
 \includegraphics[width=18cm]{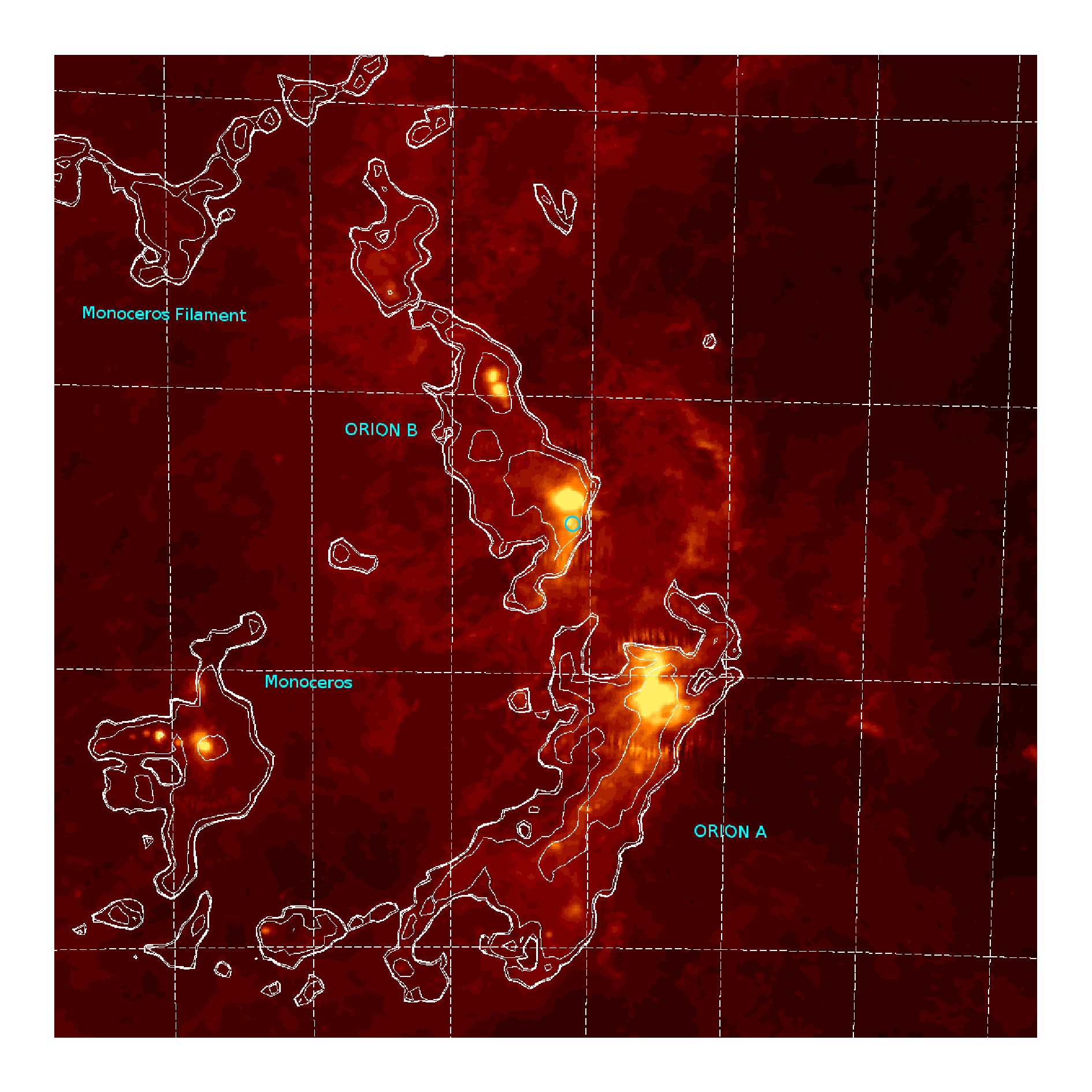}
   \caption{IRAS 100$\mu$m mosaic of the complex molecular system Orion A, Orion B 
   and Monoceros. CO map contours are overplotted as continuous lines. The light 
   open circle in the center of the image represents the position and field of view of the 
   \textit{XMM-Newton} observation.  
   }
   \label{comap}
 \end{figure*}
%_____________________________________________________________

%_____________________________________________________________
 \begin{figure*}
 \centering
 \includegraphics[]{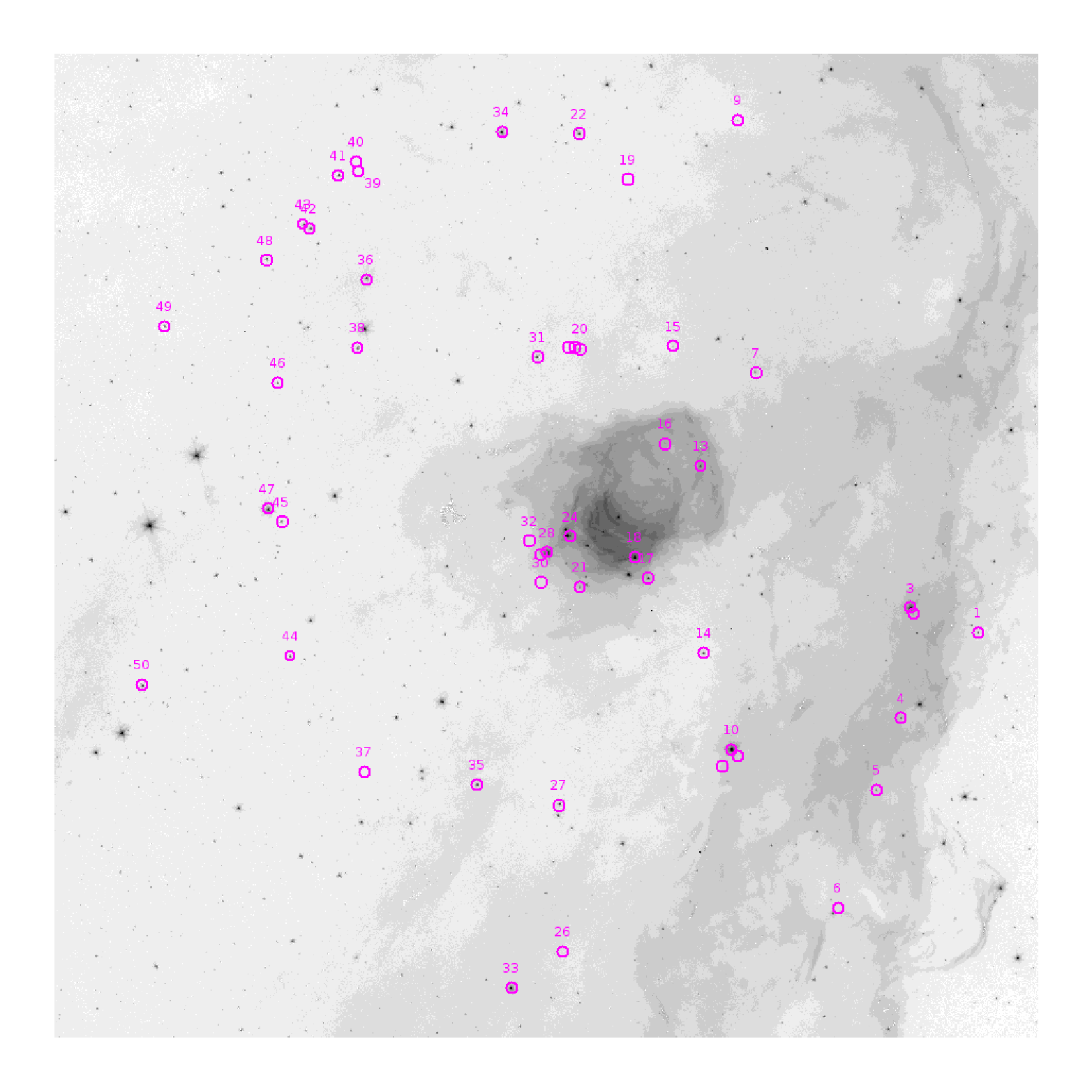}
   \caption{Spitzer-IRAC channel 1 (3.6 $\mu$m) image of the NGC 2023 nebula
   and its surroundings. The Horsehead Nebula is visible at the right, bottom corner. 
   X-ray sources detected in the \textit{XMM-Newton} observation are marked and 
   numbered in magenta. To avoid confusion, the numbers of the likely
   spurious X-ray sources identified with numbers 11 and 12 (probable 
   \textit{multi-detection} of Src 10), 23 and 25 (\textit{multi-detection} 
   of Src 20) and 29 (\textit{multi-detection} of Src 28) have been removed 
   (see Section~\ref{cross}).  
   }
   \label{fA1}
 \end{figure*}
%_____________________________________________________________

\renewcommand{\thesubfigure}{\thefigure.\arabic{subfigure}} \makeatletter
\renewcommand{\p@subfigure}{}
\renewcommand{\@thesubfigure}{\thesubfigure:\hskip\subfiglabelskip} \makeatother

%_____________________________________________________________
 \begin{figure*}
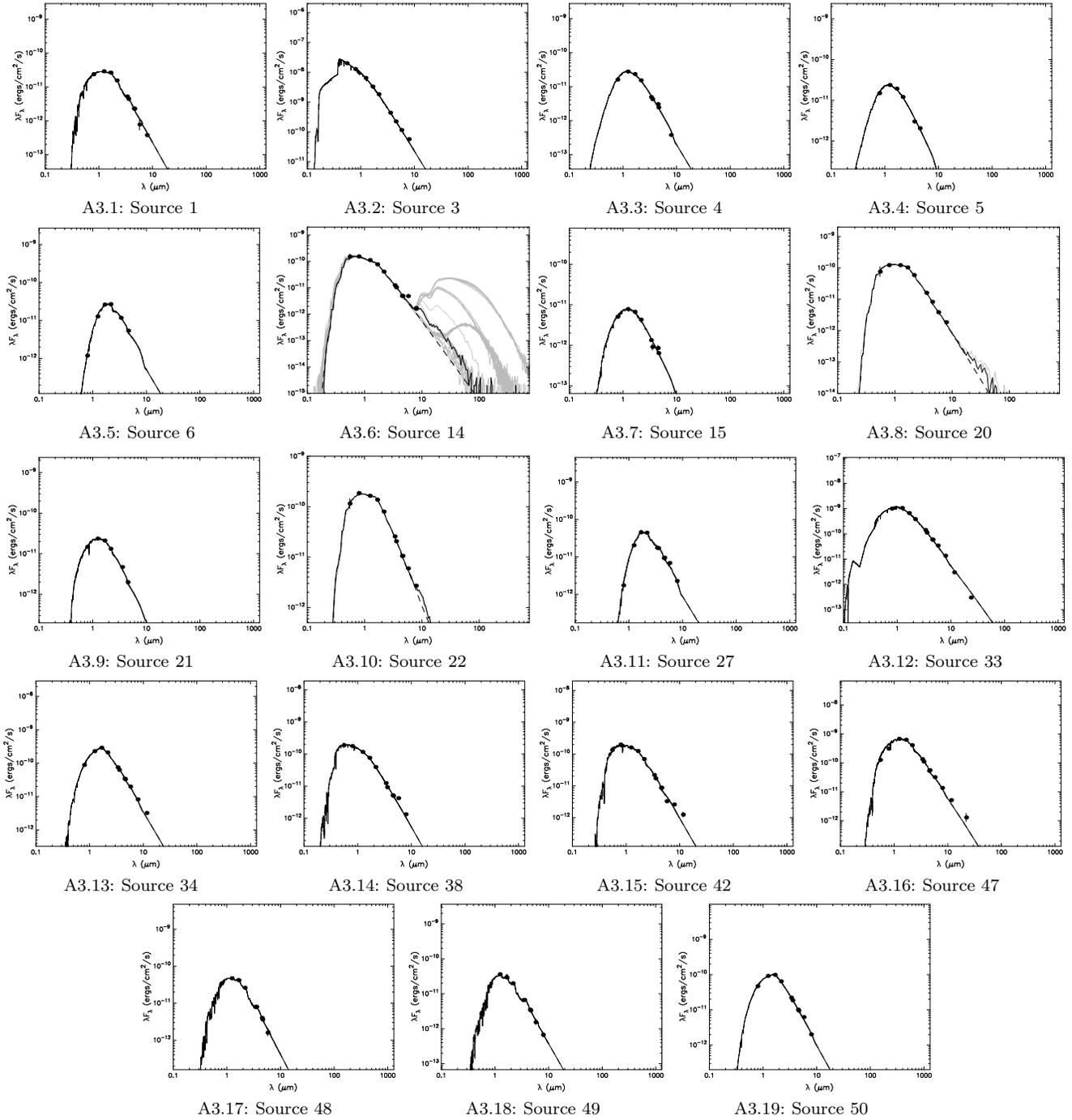

 \centering
  \subfigure[Source 1]{\includegraphics[width=4.2cm]{src1_str.eps}}
  \subfigure[Source 3]{\includegraphics[width=4.2cm]{src3.eps}}
  \subfigure[Source 4]{\includegraphics[width=4.2cm]{src4_star.eps}}
  \subfigure[Source 5]{\includegraphics[width=4.2cm]{src5_star.eps}}
  \subfigure[Source 6]{\includegraphics[width=4.2cm]{src6_star.eps} }
  \subfigure[Source 14]{\includegraphics[width=4.2cm]{src14.eps}}
  \subfigure[Source 15]{\includegraphics[width=4.2cm]{src15_star.eps}}
  \subfigure[Source 20]{ \includegraphics[width=4.2cm]{src20.eps}}
  \subfigure[Source 21]{ \includegraphics[width=4.2cm]{src21_star.eps}}
  \subfigure[Source 22]{ \includegraphics[width=4.2cm]{src22.eps}}
  \subfigure[Source 27]{ \includegraphics[width=4.2cm]{src27_star.eps}}
  \subfigure[Source 33]{ \includegraphics[width=4.2cm]{src33_star.eps}}
  \subfigure[Source 34]{\includegraphics[width=4.2cm]{src34_star.eps}}
  \subfigure[Source 38]{ \includegraphics[width=4.2cm]{src38_star.eps}}
  \subfigure[Source 42]{ \includegraphics[width=4.2cm]{src42_star.eps}}
  \subfigure[Source 47]{ \includegraphics[width=4.2cm]{src47_star.eps}}
  \subfigure[Source 48]{ \includegraphics[width=4.2cm]{src48_star.eps}}
  \subfigure[Source 49]{ \includegraphics[width=4.2cm]{src49_star.eps}}
  \subfigure[Source 50]{ \includegraphics[width=4.2cm]{src50_star.eps}}
 \caption{Spectral energy distribution of objects classified as class III in Table~\ref{tabIR}. 
                The solid black lines indicate the best-fitting model using \citet{rob07}.
                Gray lines are other models with low  $\chi^2$ values similar to that of the 
                best-fitting model ($(\chi^2 - \chi^2_\mathrm{best})/\chi^2 \le 10\%$). 
                The dashed lines represent the stellar photosphere for the accepted model.}
\label{seds_classIII}
\end{figure*}
%_____________________________________________________________

%_____________________________________________________________
 \begin{figure*}
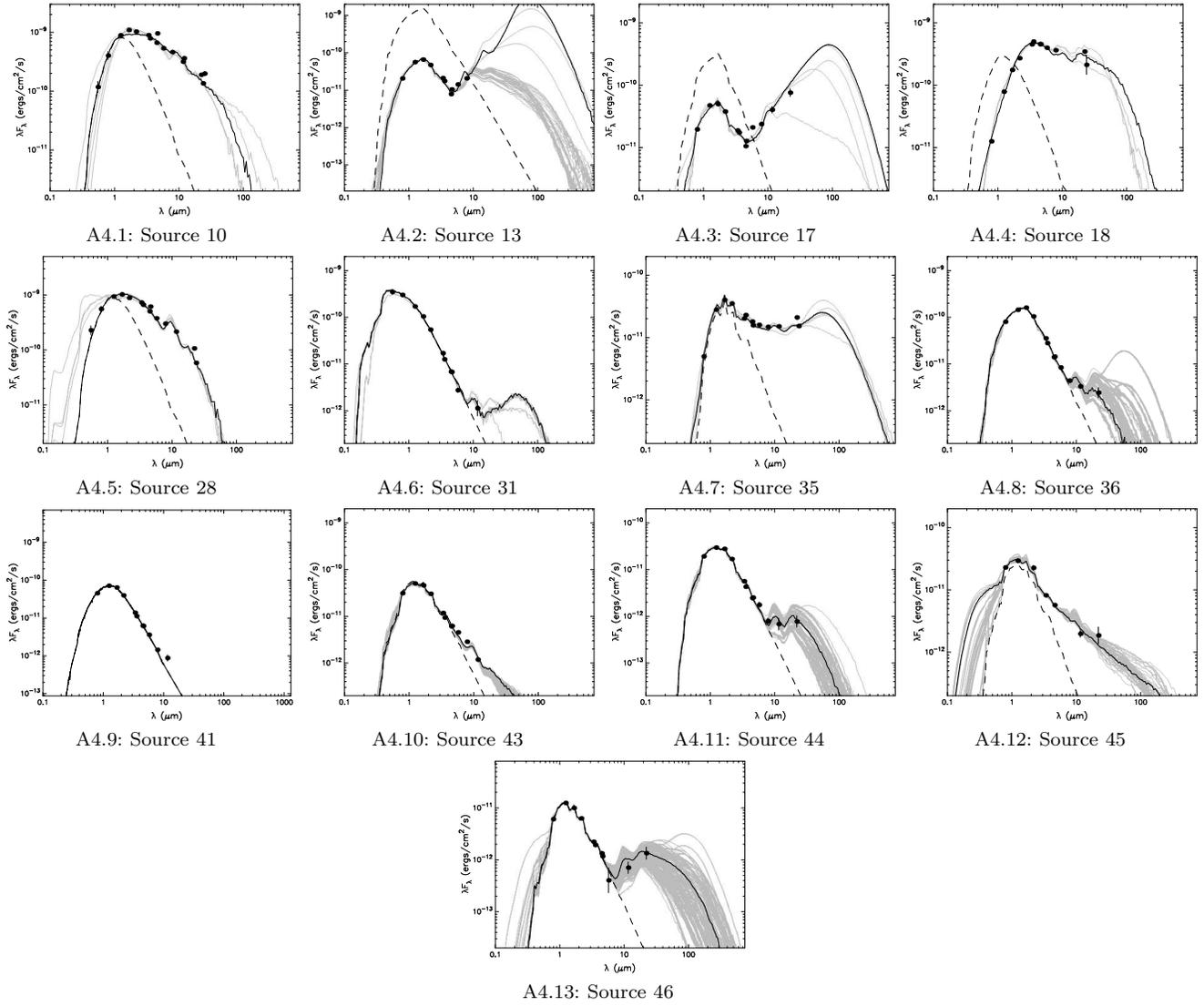

 \centering
  \subfigure[Source 10]{\includegraphics[width=4.2cm]{src10.eps}}
  \subfigure[Source 13]{\includegraphics[width=4.2cm]{src13.eps}}
  \subfigure[Source 17]{\includegraphics[width=4.2cm]{src17.eps}}
  \subfigure[Source 18]{\includegraphics[width=4.2cm]{src18.eps}}
  %\subfigure[Source 24]{\includegraphics[width=4.2cm]{src24.eps}}
  \subfigure[Source 28]{ \includegraphics[width=4.2cm]{src28.eps}}
  \subfigure[Source 31]{ \includegraphics[width=4.2cm]{src31.eps}}
  \subfigure[Source 35]{ \includegraphics[width=4.2cm]{src35.eps}}
  \subfigure[Source 36]{ \includegraphics[width=4.2cm]{src36.eps}}
  \subfigure[Source 41]{ \includegraphics[width=4.2cm]{src41_star.eps}}
  %\subfigure[Source 42]{ \includegraphics[width=4.2cm]{src42_star.eps}}
  \subfigure[Source 43]{ \includegraphics[width=4.2cm]{src43.eps}}
  \subfigure[Source 44]{ \includegraphics[width=4.2cm]{src44.eps}}
  \subfigure[Source 45]{ \includegraphics[width=4.2cm]{src45.eps}}
  \subfigure[Source 46]{ \includegraphics[width=4.2cm]{src46.eps}}
  %\subfigure[Source 47]{ \includegraphics[width=4.2cm]{src47_star.eps}}
 \caption{Same as Fig.~\ref{seds_classIII} for stars classified as class I/II, II and II/III 
 in Table~\ref{tabIR}. The dashed lines are the unabsorbed stellar SED.}
 \label{seds_classII}
 \end{figure*}
%_____________________________________________________________

\end{document}